\documentclass[a4paper,fleqn,usenatbib]{mnras}

\usepackage{newtxtext,newtxmath}

\usepackage[T1]{fontenc}
\usepackage{ae,aecompl}

\usepackage{graphicx}	
\usepackage{amsmath}	
\usepackage{amssymb}	
\usepackage{array}
\usepackage[dvipsnames]{xcolor}

\definecolor{olive}{rgb}{0, 0.7, 0}
\newcommand{\red}[1]{\textcolor{red}{#1}}
\newcommand{\green}[1]{\textcolor{olive}{#1}}

\title[Morpho-z: improving photometric redshifts]{Morpho-z: improving photometric redshifts with galaxy morphology}

\author[J. Y. H. Soo et al.]{
John Y. H. Soo$^{1}$\thanks{E-mail: yue.soo.14@ucl.ac.uk}, Bruno Moraes$^{1}$, Benjamin Joachimi$^{1}$, William Hartley$^{1}$, \and
Ofer Lahav$^{1}$, Ald\'ee Charbonnier$^{2}$, Mart\'in Makler$^{3}$, Maria E. S. Pereira$^{3}$, \and Johan Comparat$^{4}$, Thomas Erben$^{5}$, Alexie Leauthaud$^{6}$, Huanyuan Shan$^{7}$ \and and Ludovic Van Waerbeke$^{8}$.
\\
$^{1}$Department of Physics and Astronomy, University College London, Gower Street, London WC1E 6BT, UK\\
$^{2}$Observat\'{o}rio do Valongo, Universidade Federal do Rio de Janeiro, Rio de Janeiro, Brazil\\
$^{3}$Coordena\c{c}\~{a}o de Cosmologia, Astrof\'{i}sica e Intera\c{c}\~{o}es Fundamentais, Centro Brasileiro de Pesquisas F\'{i}sicas, Rio de Janeiro, Brazil\\
$^{4}$Max-Planck-Institut f\"{u}r extraterrestrische Physik (MPE), Giessenbachstrasse 1, D-85748 Garching bei M\"unchen, Germany\\
$^{5}$Argelander-Institut f\"ur Astronomie, Auf dem H\"ugel 71, 53121 Bonn, Germany\\
$^{6}$Department of Astronomy and Astrophysics, University of California, Santa Cruz, 1156 High Street, Santa Cruz, CA 95064, USA\\
$^{7}$Laboratoire d'Astrophysique, Ecole Polytechnique F\'ed\'erale de Lausanne (EPFL), Observatoire de Sauverny, CH-1290 Versoix, Switzerland\\
$^{8}$Department of Physics and Astronomy, University of British Columbia, 6224 Agricultural Road, Vancouver, B.C., V6T 1Z1, Canada\\
}

\date{Accepted XXX. Received YYY; in original form ZZZ}

\pubyear{2017}

\begin{document}
\label{firstpage}
\pagerange{\pageref{firstpage}--\pageref{lastpage}}
\maketitle

\begin{abstract}
We conduct a comprehensive study of the effects of incorporating galaxy morphology information in photometric redshift estimation. Using machine learning methods, we assess the changes in the scatter and outlier fraction of photometric redshifts when galaxy size, ellipticity, S\'{e}rsic index and surface brightness are included in training on galaxy samples from the SDSS and the CFHT Stripe-82 Survey (CS82). We show that by adding galaxy morphological parameters to full $ugriz$ photometry, only mild improvements are obtained, while the gains are substantial in cases where fewer passbands are available. For instance, the combination of $grz$ photometry and morphological parameters almost fully recovers the metrics of $5$-band photometric redshifts. We demonstrate that with morphology it is possible to determine useful redshift distribution $N(z)$ of galaxy samples without any colour information. We also find that the inclusion of quasar redshifts and associated object sizes in training improves the quality of photometric redshift catalogues, compensating for the lack of a good star-galaxy separator. We further show that morphological information can mitigate biases and scatter due to bad photometry. As an application, we derive both point estimates and posterior distributions of redshifts for the official CS82 catalogue, training on morphology and SDSS Stripe-82 $ugriz$ bands when available. Our redshifts yield a 68th percentile error of $0.058(1+z)$, and a outlier fraction of $5.2$ per cent. We further include a deep extension trained on morphology and single $i$-band CS82 photometry.

\end{abstract}

\begin{keywords}
galaxies: distances and redshifts -- galaxies: structure -- methods: statistical
\end{keywords}

\section{Introduction}\label{sec:intro}
Redshifts of galaxies provide distance information for many cosmological analyses, and are especially needed to study large scale structure. While the most accurate way to estimate redshifts is through spectroscopy, it is unfortunately a very expensive and time-consuming process. Thus, in order to produce redshifts for ideally all objects in large galaxy samples, high-quality photometric redshifts (photo-$z$'s) are much sought after, for example in weak lensing studies where there is need for accurate and unbiased knowledge of the redshift distribution $N(z)$ for ensemble samples. photo-$z$'s are typically estimated using broadband magnitudes, obtained using two main approaches. The first approach uses spectral template fitting, e.g. codes like \textsc{le phare} \citep{arnouts_measuring_1999}, \textsc{bpz} \citep{benitez_bayesian_2000}, \textsc{hyperz} \citep{bolzonella_photometric_2000}, \textsc{zebra} \citep{feldmann_zurich_2006}, \textsc{eazy} \citep{brammer_eazy:_2008}, \textsc{gazelle} \citep{kotulla_impact_2009} and \textsc{delight} \citep{leistedt_data-driven_2017}. The second approach uses empirical / machine learning techniques, e.g. artificial neural networks \citep{firth_estimating_2003,collister_annz:_2004,sadeh_annz2:_2016}, multi-layered perceptron \citep{vanzella_photometric_2004,brescia_catalogue_2014}, support vector machines \citep{wadadekar_estimating_2005}, Gaussian process regression \citep{way_novel_2006}, boosted decision trees \citep{gerdes_arborz:_2010}, random forests \citep{carrasco_kind_tpz:_2013,rau_accurate_2015}, genetic algorithms \citep{hogan_gaz:_2015} and sparse  Gaussian framework \citep{almosallam_sparse_2016}.

With the availability of such a wide range of photo-$z$ codes and methods, comparisons of various implementations have been performed \citep{hildebrandt_phat:_2010,abdalla_comparison_2011,sanchez_photometric_2014}. No obvious best photo-$z$ code was named since each code displays different strengths depending on the metrics used. The focus of recent photo-$z$ analyses has turned to improving error estimation \citep{oyaizu_photometric_2008,hoyle_anomaly_2015,wittman_overconfidence_2016}, the use of new statistical techniques \citep{lima_estimating_2008,zitlau_stacking_2016}, improving existing algorithms \citep{cavuoti_photometric_2015,sadeh_annz2:_2016}, and the addition of extra input information to get more precise and accurate photo-$z$'s. With regards to the inclusion of extra information, a recent example in template methods includes using surface brightness as a prior in spectral energy distribution (SED) templates \citep{kurtz_-photoz:_2007,stabenau_photometric_2008}, motivated by the knowledge of surface brightness dimming $(1+z)^4$. In empirical methods this application is more straightforward, since algorithms are constructed such that it is not difficult to add extra input parameters. For example, \citet{collister_annz:_2004} and \citet{wadadekar_estimating_2005} demonstrated that by including the $50$ and $90$ per cent Petrosian flux radii ($R_{\rm P50}$, $R_{\rm P90}$), the photometric redshift root-mean-square errors improve by $3$ and $15$ per cent respectively for the SDSS main galaxy sample. \citet{tagliaferri_neural_2003} used Petrosian fluxes and radii in their work on galaxies from the SDSS early data release and calculated robust errors decreasing as much as 24 per cent. Meanwhile, \citet{vince_toward_2007} included the concentration of galaxy light profiles in their study and reported that the root-mean-square error of photo-$z$'s on SDSS galaxies improved by 3 per cent. \citet{wray_new_2008} included surface brightness and the S\'{e}rsic index, and found improvements in variance when compared to other template fitting methods applied to the SDSS main galaxy sample previously.

A particularly thorough investigation was performed by \citet{way_new_2009}, who studied how galaxy morphology information affects photometric redshift quality. Using the Gaussian process regression method, they included several galaxy morphological parameters alongside with photometry for training, like $R_{\rm P50}$, $R_{\rm P90}$, concentration index ($C$, the ratio between the two Petrosian flux radii), \texttt{fracDeV} (the weight of the de Vaucouleurs component in the best composite model) and the Stokes $Q$ parameter (a measurement of ellipticity). They showed that the addition of these parameters does not systematically improve the photometric redshift estimation. Later, \citet{way_galaxy_2011} separated galaxy samples into ellipticals and spirals with the help of Galaxy Zoo \citep{lintott_galaxy_2011}, and showed that photometric redshifts estimated using adaptive moments and texture for the SDSS luminous red galaxies yield a root-mean-square error as low as $0.012$. 

\citet{singal_efficacy_2011} formed principal components of a series of 8 derived morphological shape parameters (including smoothness, asymmetry and Gini coefficient) and used these in combination with photometry to improve photometric redshift estimations. However, they found that outliers were not significantly decreased, and the shape parameters may have contributed noise instead. \citet{jones_analysis_2017} repeated the study with their support vector machine code \textsc{spiderz}, and obtained results in agreement with the earlier work.

It is natural to expect that different surveys and redshift ranges will benefit to different degrees from galaxy morphology information, however due to different results reported by different groups, a comprehensive study on this subject is warranted. Many current and upcoming surveys such as the Hyper-Suprime Cam\footnote{\url{http://www.naoj.org/Projects/HSC/}} (HSC), Kilo-Degree Survey\footnote{\url{http://kids.strw.leidenuniv.nl/}} (KiDS), Dark Energy Survey\footnote{\url{http://www.darkenergysurvey.org/}} (DES), Large Synoptic Survey Telescope\footnote{\url{http://www.lsst.org/}} (LSST), Wide Field Infrared Survey Telescope\footnote{\url{https://wfirst.gsfc.nasa.gov/}} (WFIRST) and Euclid\footnote{\url{http://sci.esa.int/euclid}} \citep{laureijs_euclid_2011} could have their photo-$z$ estimation methods benefit from the high fidelity galaxy morphological parameters that would come for free.

Therefore, this work is aimed at studying the extent to which different morphological parameters improve photometric redshifts, what kind of objects or photometric conditions benefit most from it, and the implications and applications of this knowledge to current and future surveys. This study is different from previous studies in that it is conducted in a more comprehensive manner: using data sets with high quality morphology, varying the number of filters, using photometry of different qualities, testing individual morphological parameters and considering the inclusion of quasar spectra. This paper will address the following questions:

\begin{enumerate}
\item Which morphological parameters improve photo-$z$'s the most in a general galaxy sample?
\item What are the impacts of morphology on photo-$z$'s in
\begin{enumerate}
\item a sample with low-quality photometry?
\item a survey with fewer than $5$ broadband filters?
\item a sample contaminated by quasars?
\end{enumerate}
\item What are the impacts of galaxy morphology on the accuracy of individual redshift probability densities (pdf henceforth) and the redshift distribution $N(z)$ of galaxies?
\end{enumerate}

This paper is structured as follows. In Section~\ref{sec:algor} we will introduce the algorithms used in this study, followed by Section~\ref{sec:data} which explains the various datasets and training samples used. Section~\ref{sec:method} discusses the morphological parameters and metrics used. Answers and discussions for question (i) are presented in Section~\ref{sec:res_general}, followed by the discussion of question (ii) in Section~\ref{sec:res_less_optimal}, and question (iii) addressed in Section~\ref{sec:res_pz}. The key results obtained are applied to the CS82 survey in Section~\ref{sec:app}, in which a photo-$z$ catalogue is produced and made available to the public. The paper is concluded in Section~\ref{sec:disc}. In this paper we define the \textit{testing set} as the set of data where the metrics of performance are evaluated; the \textit{validation set} is used as part of the training process to prevent overtraining; while the \textit{target set} is the set of data where no redshift information is available and where photo-$z$'s are estimated.

\section{Algorithms}\label{sec:algor}
In this work we employ machine learning techniques to estimate photo-$z$'s. These machine learning algorithms find a deterministic relationship between the input variables (e.g. $ugriz$ broadband magnitudes) and the spectroscopic redshifts in a training set (where we assume that the spectroscopic redshift is the true redshift), and this information is then used to produce photo-$z$'s for a target sample for which no spectroscopic redshift information is available.

In the following paragraphs we briefly summarise the two algorithms: \textsc{annz} and \textsc{annz2} used in this work. 

\subsection{ANNz}\label{sec:annz}
\textsc{annz}\footnote{\url{http://www.homepages.ucl.ac.uk/~ucapola/annz.html}} is an artificial neural network (ANN) redshift estimation library introduced by \citet{collister_annz:_2004}. An ANN is made up of interconnected nodes arranged in several layers. In a typical setup, the input nodes are the broadband magnitudes, which are then processed in the inner layers, and the single output node is the photo-$z$. Each connection between nodes carries a weight, and each node carries a value which is calculated via the summation of weights and activation functions (usually sigmoids, making the process highly non-linear) of nodes connected to it from the previous layer. The network will constantly adjust the weights such that a cost function (here the difference between the spectroscopic and photometric redshift) is minimized. The training objects are divided into a training and validation set to prevent over-fitting. Readers can refer to \citet{collister_annz:_2004} for more details.

\textsc{annz} is only used in Section~\ref{sec:res_quasar} to obtain photo-$z$ point estimates for the Stripe-82 sample. This is done to ensure a fair comparison with the photo-$z$ produced by \citet{reis_sloan_2012}, since both codes are very similar to one another. For each training in this study, $4$ committees of networks (i.e. $4$ training rounds each with a different random number) are used, using an architecture of N:2N:2N:1 (N inputs, $2$ hidden layers with 2N nodes each, and one output). This is the default setting for \textsc{annz}, we have tested that increasing the number of hidden layers and nodes do not have significant effect on the performance of training.

\subsection{ANNz2}\label{sec:annz2}
Despite similar names, \textsc{annz2}\footnote{\url{https://github.com/IftachSadeh/ANNZ}} \citep{sadeh_annz2:_2016} is independent of \textsc{annz}, and they differ in programming language and functionality. \textsc{annz2} uses the Toolkit for Multivariate Data Analysis (TMVA) with \textsc{root}\footnote{\url{http://tmva.sourceforge.net/}}, which is a powerful package that incorporates several machine learning algorithms including ANN, boosted decision trees (BDT) and k-nearest neighbours (KNN). \textsc{annz2} can run multiple machine learning algorithms for a single training and outputs photo-$z$'s based on a weighted average of their performances.

Other than the usual regression method to obtain point estimates for photo-$z$'s, \textsc{annz2} is also able to produce redshift posterior probability distributions $P(z)$, conduct classification and support reweighting between samples. photo-$z$'s in the form of pdfs have been shown to produce superior results compared to point estimates in weak lensing and clustering measurements \citep{mandelbaum_precision_2008,myers_incorporating_2009,benjamin_cfhtlens_2013,sanchez_photometric_2014,kuijken_gravitational_2015,jouvel_photometric_2017}. These pdfs are produced by propagating the intrinsic uncertainty on the input parameters and the uncertainty in the machine learning method to the expected photo-$z$ solution. 

\textsc{annz2} also differs from \textsc{annz} in its presentation of photo-$z$ uncertainties. \textsc{annz} derives uncertainties through error propagation of photometric errors from the inputs and network variance of the neural network, while \textsc{annz2} derives uncertainties using the KNN method: first it estimates the photo-$z$ bias between each object and a fixed number of nearest neighbours in parameter space, it then takes the 68th percentile width of the distribution of the bias. This asserts that objects with similar photometric properties should have similar uncertainties, and this error presentation has been shown to perform better than the former \citep{sadeh_annz2:_2016}.

\textsc{annz2} produces four point estimates for each run, only two of which are used in this work. The first is the peak photo-$z$ $z_{\rm peak}$, which is the position of the highest peak\footnote{Due to the possibility of multiple peaks in a photo-$z$ pdf, the word `peak' from this point onwards refers to the highest peak in the pdf distribution, unless stated otherwise.} of the pdf generated. As the values of pdfs are produced in bins of 0.01 in redshift, and the appearance of noisy pdfs for high redshift objects, $z_{\rm peak}$ may not be the most accurate photo-$z$ point estimate to use, but it is used to determine the ODDS value of an object (see Section~\ref{sec:odds}). The other is the pdf average photo-$z$ (denoted $z_{\rm phot}$ throughout this paper), which is the mean photo-$z$ calculated over the pdf. The pdf average photo-$z$ will act as the best point estimate photo-$z$ output for \textsc{annz2} throughout this paper.

\textsc{annz2} has been widely used in recent work \citep{sanchez_photometric_2014,bonnett_using_2015,jouvel_photometric_2017}; it was selected as the primary algorithm in this study due to its high customizability and its ability to produce pdfs. \textsc{annz2} is used in Sections~\ref{sec:res_general} through~\ref{sec:app} to obtain $z_{\rm phot}$ and pdfs for each galaxy, and $N(z)$ of the distributions. In this study, a set of 5 ANNs with different random seeds are used during each training. Only ANNs are used for consistency, it also allows fair comparison between trainings with different numbers of inputs. The ANN settings and architecture used for \textsc{annz2} are similar to those used in \textsc{annz}.

\subsection{Reweighting algorithm}\label{sec:reweight}
Spectroscopic galaxy samples are usually constructed by cross-matching photometry from large photometric surveys with redshifts obtained from multiple different spectroscopic surveys. Due to the different target selections of each spectroscopic survey, the combined spectroscopic training sample will contain objects which are unevenly distributed in colour-magnitude space, e.g. contain preferentially bright and red galaxies. This means that the distribution of training parameters (e.g. magnitude, colour, size etc.) in the training, validation and testing samples will turn out to be quite different from that of the target sample, in which the photo-$z$'s are to be estimated. Therefore reweighting of the spectroscopic sample to become representative of the training parameters of the target sample is needed, not only to ensure that the metrics evaluated on the testing set is representative of the target set, but also to ensure that none of the spectroscopic sources are over-represented in the training (see Section~\ref{sec:spectroscopy} for more details).

In this study we adopt the reweighting method as introduced by \citet{lima_estimating_2008}, which is done by comparing the density of objects in a selected parameter space of the spectroscopic and target samples, and setting a weight value to each object in the spectroscopic sample so that during the training process, the cost function used to estimate the photo-$z$ will be balanced by upweighting objects that are less represented in the training sample compared to the target sample, and downweighting objects otherwise. Since the testing set (where performance metrics are evaluated) is also drawn from the spectroscopic sample, the weights have to be taken into account when estimating the metrics to reflect the photo-$z$ performance of the target sample \citep[See][for more details]{sanchez_photometric_2014}.

\textsc{annz2} allows weights to be incorporated during the training process, however the in-built reweighting code in \textsc{annz2} is not used in this project. We use an external reweighting algorithm similar to one used in \citet{sanchez_photometric_2014} to calculate the individual weights of the objects. This algorithm first uses a k-dimensional tree to bin the objects in the parameter space assigned. It then proceeds to calculate the number of nearest neighbours of each object in the sample. The weights are then derived by calculating the ratio of the densities between the training and the target sample. The weights obtained are used in \textsc{annz2} to calculate the photo-$z$'s.

In this study we reweight the objects in only the $i$-band and $g$-$i$ colour, or $i$-band and radius $r_{\rm exp}$ when colour is not available. We do not use more than $2$ parameters as the reweighting algorithm would overfit and create biases. It was also found that the other input parameters are well reweighted by just using these two parameters alone, see Section~\ref{sec:param} for more details. Since we intend to evaluate the impact of morphology on photo-$z$'s not on the spectroscopic sample but instead on a sample representative of current and future surveys, objects in all spectroscopic samples used in this paper are weighted with respect to the CS82 target sample, the only exception being Section~\ref{sec:res_quasar}, since the metrics calculated for the Stripe-82 extended objects sample are used in direct comparison with results from \citet{reis_sloan_2012}.

\section{Data}\label{sec:data}
Multi-band photometric, high resolution imaging and spectroscopic galaxy data are needed in order to study how galaxy morphology affects the quality of photometric redshifts. Sections~\ref{sec:photometry} and \ref{sec:spectroscopy} describe the sources of photometric (broadband magnitudes and galaxy morphology) and spectroscopic (redshift) data for this study, while Section~\ref{sec:samples} describes all the training samples used in this paper.

\subsection{Photometry}\label{sec:photometry}

\subsubsection{SDSS Stripe-82 Survey}\label{sec:stripe82}
The SDSS Stripe-82 Survey \citep{annis_sloan_2014} is a co-addition of the SDSS Stripe-82 imaging data \citep{jiang_sloan_2014}, obtained by repeated scanning along the stripe at the equator between $-50^{\circ}\leq$RA$\leq 60^{\circ}$ and $-1.25^{\circ}\leq$DEC$\leq 1.25^{\circ}$, reaching about $2$ magnitudes fainter than standard SDSS observations ($i\sim 24.1$). We chose to work with objects in the Stripe-82 region due to the availability of photometric data from SDSS, galaxy morphology from CS82 (see Section~\ref{sec:cs82}) and the abundance of spectroscopic redshifts in this region. These wide-angle deep imaging data have been used extensively in many projects, e.g. photo-$z$ computation \citep{reis_sloan_2012}, quasar classification \citep{peters_quasar_2015}, massive galaxy evolution \citep{bundy_stripe_2015} and deeper co-adds \citep{fliri_iac_2016}. The co-added photometric data in this region can be obtained from the SDSS CasJobs\footnote{\url{http://skyserver.sdss.org/CasJobs/}}, setting \texttt{run=106,206}. Runs other than these are either photometric data prior to co-addition, or were data not observed under photometric conditions. The number of galaxies and quasars and the selection cuts used are discussed in Section~\ref{sec:samples}.

\subsubsection{CFHT Stripe-82 Survey (CS82) morphology catalogue} \label{sec:cs82}
The CS82 Survey is a joint Canada-France-Brazil project. Using the MegaCam at the Canada France Hawaii Telescope (CFHT), it surveyed approximately $170$ deg$^2$ of the equatorial Stripe-82 area. It is a relatively deep survey that maps down to magnitude $24.1$ in the $i$-band (which is the only band in this survey), and has a mean seeing of $0\farcs 6$. Data from this survey has been used for several weak lensing analyses \citep[e.g.][Pereira, et al. in prep.; Vitorelli et al. in prep.]{comparat_stochastic_2013,shan_weak_2014,hand_first_2015,liu_cosmological_2015,battaglia_weak-lensing_2016,li_measuring_2016,niemiec_stellar--halo_2017,leauthaud_lensing_2017,shan_mass-concentration_2017}. A galaxy morphology catalogue has been produced from CS82 data \citep[][Moraes et al. in prep.]{charbonnier_abundance_2017} using \textsc{sextractor}\footnote{\url{https://www.astromatic.net/software/sextractor}} and \textsc{psfex}\footnote{\url{https://www.astromatic.net/software/psfex}} to fit a series of single-component profiles (de Vaucouleurs, exponential and S\'ersic). The galaxy morphological parameters taken from this catalogue are discussed in Section~\ref{sec:param}.

\subsection{Spectroscopy}\label{sec:spectroscopy}
\begin{figure*}   
\includegraphics[width=0.75\linewidth]{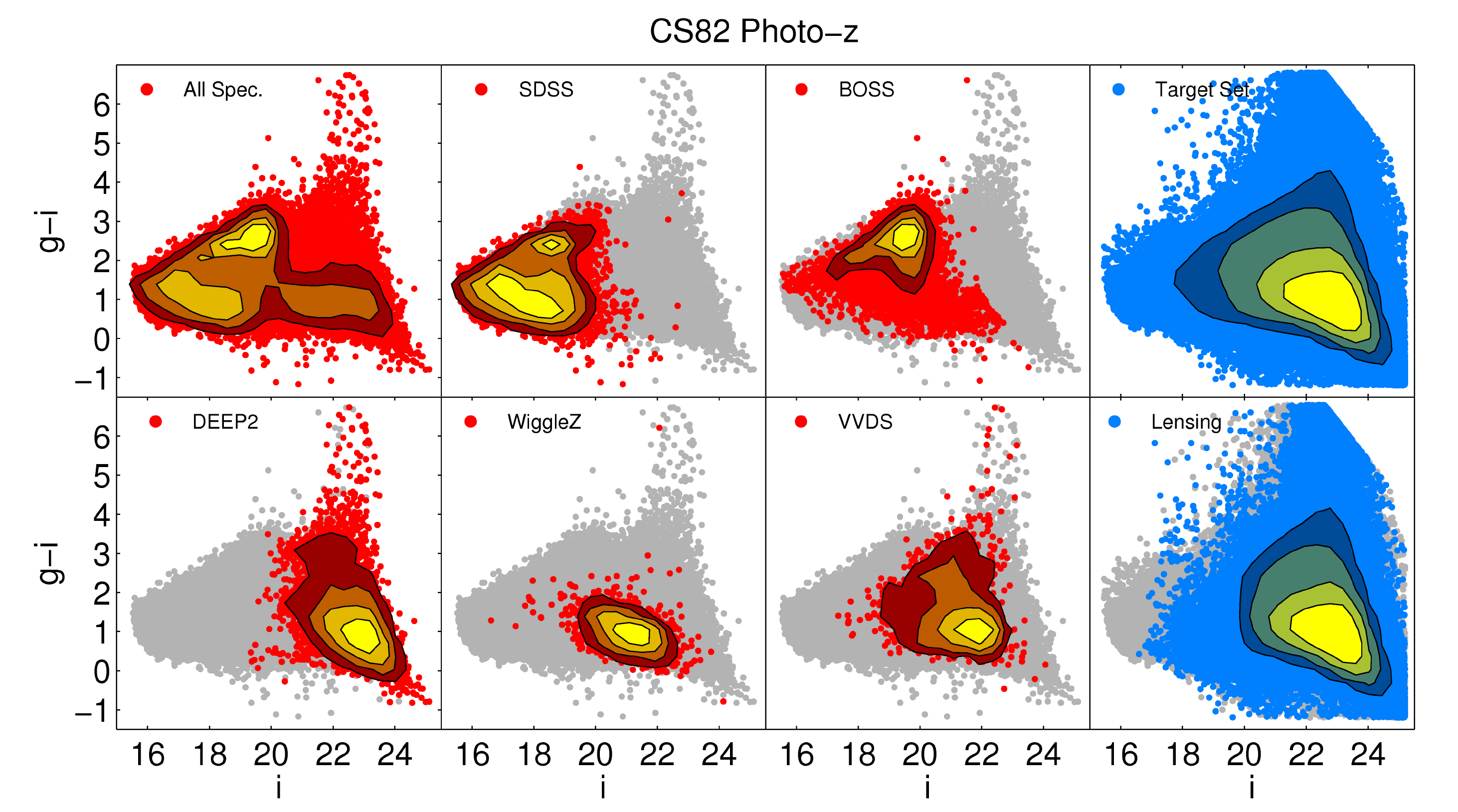}
\caption{Apparent $g$-$i$ \textit{vs} $i$ magnitude for objects from the CS82 training sample (red), and part of the CS82 target sample with matched SDSS photometry (blue) used in this study. The contours reflect the density of the objects. The six panels on the left highlight the training objects according to their spectroscopic sources (SDSS, BOSS, DEEP2, WiggleZ and VVDS), while the bottom right plot highlights the lensing subsample.} \label{fig:corr_spec}
\end{figure*}

As empirical photo-$z$ methods require the use of training samples containing true redshifts, all photometric and morphological samples mentioned previously will have to be cross-matched with spectroscopic data as outlined below to form the data samples needed in this study. Spectroscopic redshifts from five surveys have been used in this study, and the distribution of these objects in colour-magnitude space is illustrated in Fig.~\ref{fig:corr_spec}.

\subsubsection{SDSS and BOSS spectroscopy}\label{sec:sdss_boss}
Spectroscopy obtained from SDSS is taken using either the SDSS or Baryon Oscillation Spectroscopic Survey (BOSS) spectrographs. The SDSS spectrograph was used in the SDSS Legacy Survey \citep{york_sloan_2000} which obtained spectroscopic redshifts for about $930\,000$ galaxies as faint as Petrosian magnitudes $r\sim 17.77$ and about 120~000 quasars up to PSF magnitudes of $i\sim 19.1$. BOSS \citep{dawson_baryon_2013} on the other hand measured redshifts of galaxies to $i\sim 19.9$, and of quasars to $g\sim 22.0$, and obtained spectroscopic redshifts for over $1.5$~million galaxies up to $z\sim 0.7$, and over $160\,000$ quasars with $2.2<z<3.0$. 

Our SDSS and BOSS spectroscopic sample selection is simple as we use all available high quality SDSS spectroscopic redshifts within the Stripe-82 region. To ensure that only good quality redshifts are used, we select all redshifts from SDSS and BOSS that have \texttt{zWarning=0}. This selection provides $75\,229$ galaxy and $5\,380$ quasar redshifts from SDSS, while BOSS provides good quality redshifts for $18\,546$ galaxies and $552$ quasars. These redshifts are cross-matched to their photometric counterparts in Stripe-82 from the \texttt{PhotoPrimary} table for $ugriz$ magnitudes and morphology.

\subsubsection{DEEP2 Redshift Survey}\label{sec:deep2}
The DEEP2 redshift survey \citep{newman_deep2_2013} used the Keck telescope to study the properties of massive galaxies and large scale structure. This is an untargeted survey which reaches a depth of $r\sim 24.1$ and covers an overlap area of $0.5$ deg$^2$ with the Stripe-82 region. This survey uses the colours $B-R$ and $R-I$ to remove objects with $z_{\rm spec}<0.7$, providing a sample of objects with $z_{\rm spec}\sim 1$. Objects from the DEEP2 DR4 redshift catalogue\footnote{\url{http://deep.ps.uci.edu/DR4/zcatalog.html}} with \texttt{ZQUALITY}$\geq 3$ were obtained and cross-matched with SDSS Stripe-82 photometry to yield $11\,858$ redshifts for this study.

\subsubsection{WiggleZ Dark Energy Survey}\label{sec:wigglez}
The WiggleZ dark energy survey \citep{drinkwater_wigglez_2010} used the Anglo-Australian Telescope to study emission-line galaxies. Its aim was to measure the precise scale of the baryon acoustic oscillation (BAO) imprinted on the spatial distribution. This survey is flux-limited in the Galaxy Evolution Explorer (GALEX) ultraviolet band $NUV<22.8$ and also detected optically at $20.0<r<22.5$, and it has a $45.3$ deg$^2$ survey area overlapped with the Stripe-82 region. Galaxy redshifts from the WiggleZ DR1 database\footnote{\url{http://wigglez.swin.edu.au/site/data.html}} with quality flag \texttt{qop}$>3$ are used, providing $7\,648$ redshifts for this study.

\subsubsection{VIMOS VLT Deep Survey (VVDS)}\label{sec:vvds}
VVDS \citep{le_fevre_vimos_2013} is a deep representative galaxy survey, which uses the VIMOS multi-slit spectrograph at the ESO-VLT. The VVDS-Wide \citep{garilli_vimos_2008} survey has an overlapping area of $3.6$~deg$^2$ within the Stripe-82 region, with a limiting magnitude of $i\sim 22.5$. It aimed to trace the large-scale distribution of galaxies up to $z\sim 1$ on comoving scales reaching $100$ h$^{-1}$ Mpc. Objects from the VVDS-Wide (VVDS-F2217+00) sample\footnote{\url{http://cesam.lam.fr/vvds/vvds_download.php}} with \texttt{zflag=3,4} (galaxies) and \texttt{zflag=13,14} (quasars) are cross-matched with the Stripe-82 sample, contributing $3\,949$ redshifts for the Stripe-82 training sample.

\subsection{Training samples used}\label{sec:samples}
The available spectroscopic data are formed into four samples for use in this study, the description and purpose of which are discussed in the paragraphs below. The first three samples are used to answer the main questions stated in Section~\ref{sec:intro}, while the last training sample is used to produce photo-$z$'s for the CS82 catalogue. The four training samples used in this study are summarised in Table~\ref{tab:sample}. Here we also remind the reader that we define `testing set' as the sample in which metrics are evaluated on, while `target set' refers to the final sample in which photo-$z$'s are evaluated. 

\begin{table*}  
\caption{Spectroscopic and photometric samples used in this study, listed with their respective sample sizes, sources of photometry, morphology and spectroscopy. Listed are also the respective sections in which these samples are featured.} \label{tab:sample}
\begin{tabular}{lrllll}
\hline
\bf Sample 					& \bf Size 		& \bf Photometry 	& \bf Morphology& \bf Spectroscopy 						& \bf Sections \\
\hline
CS82 general sample 		& $59\,498$		& S82 Co-add		& CS82 			& SDSS, BOSS, DEEP2, WiggleZ, VVDS		& \ref{sec:res_diff_param}, \ref{sec:res_less_filter},  \ref{sec:res_pz} \\
S82 low-quality sample		& $57\,784$	 	& S82				& CS82 			& SDSS, BOSS, DEEP2, WiggleZ, VVDS		& \ref{sec:res_bad_photometry} \\
S82 extended objects		& $97\,812$ 	& S82 Co-add		& S82 Co-add	& SDSS, DEEP2, WiggleZ, VVDS			& \ref{sec:res_quasar}\\
CS82 photo-$z$ training set	& $64\,591$		& S82 Co-add, CS82	& CS82 			& SDSS, BOSS, DEEP2, WiggleZ, VVDS		& \ref{sec:app}	\\
\hline
CS82 photo-$z$ target set		& $5\,777\,379$	& S82 Co-add, CS82	& CS82 			& - 									& \ref{sec:app}	            \\
~~Lensing subset 			& $3\,536\,783$	& S82 Co-add, CS82 	& CS82			& - 									& \ref{sec:app}				\\
\hline
\end{tabular}
\end{table*}

The first spectroscopic sample is denoted as the CS82 general sample. This sample uses $ugriz$ photometry from SDSS Stripe-82 Co-add, morphology from the CS82 morphology catalogue, and spectroscopic redshifts from SDSS, BOSS, DEEP2, WiggleZ and VVDS. The selection criteria for this sample are as follows. For SDSS photometry, only objects from \texttt{run=106,206} (co-added photometry) and magnitudes within $16.0<r<24.5$ are used; for CS82 morphology, we require \texttt{MASK=0} (not masked), \texttt{0$\leq$FLAGS$\leq$3} (flag for good quality source extraction), \texttt{MAGERR\_AUTO<0.1086} (S/N ratio $>10$) and \texttt{SPREAD\_MODEL\_SER>0.008} (S\'ersic spread model, a star-galaxy separator to select only extended objects, justification discussed in Moraes et al. in prep.); and finally for spectroscopy, we require that they have spectral \texttt{class} set as \texttt{GALAXY} in addition to the quality cuts for each source mentioned in Section~\ref{sec:spectroscopy}. Small selection effects may arise since objects with bad galaxy morphology data (radius, axis-ratio and other parameters used in this study are discussed in Section~\ref{sec:param} below) from both SDSS and CS82 have to be removed. The selection cuts above produce a sample of $59\,498$ galaxies. This sample is used in three sections: the first in Section~\ref{sec:res_diff_param} to study the effects of individual or multiple morphological parameters on photo-$z$'s; next in Section~\ref{sec:res_less_filter} this sample is used to study the effects of galaxy morphology on surveys with limited number of broadband filters; and lastly in Section~\ref{sec:res_pz} this sample is used to study the effects of galaxy morphology on galaxy pdfs and redshift distributions.

The second spectroscopic sample is the Stripe-82 low-quality photometry sample, which is very similar to the previous sample but uses lower quality photometry than the former. This set uses the same cuts and sources for spectroscopy and morphology, but cross-matched with photometry from SDSS Stripe-82 not necessarily from runs $106$ and $206$, but still ensuring that these were \texttt{PRIMARY} objects (see Section~\ref{sec:res_bad_photometry} for more information). In other words, this sample contains a mixture of high-quality co-added photometry and low-quality photometry which was taken under non-photometric conditions. We used this sample to study the impact of galaxy morphology on photo-$z$'s from surveys which lack good quality photometry in Section~\ref{sec:res_bad_photometry}. This sample yields  $57\,784$ galaxies, of which $58$ per cent of the objects have low-quality photometry. There are more objects with invalid / erroneous morphological data in this sample than the former, thus the removal of these objects resulted in a slightly smaller sample.

The third spectroscopic sample is the Stripe-82 extended objects sample. This sample differs from the previous samples in that it does not use morphology from the CS82 Survey and spectroscopic redshifts from BOSS, as we intend to use a spectroscopic sample as close as possible to the sample choice of \citet{reis_sloan_2012}. Photometry from SDSS Stripe-82 runs $106$ and $206$ is cross-matched with galaxy and quasar redshifts from SDSS, DEEP2, WiggleZ and VVDS to form a sample with $104\,286$ objects. In this sample we used the SDSS star-galaxy separator \texttt{type=3} to keep only extended objects. As the star-galaxy separator is not perfect, we expect to see a mixture of galaxies and quasars in the sample. This sample yields $97\,812$ objects in which only $97$ per cent have galaxy spectra. In Section~\ref{sec:res_quasar} we use this sample to study whether the photo-$z$ of a group of extended objects improve when quasar redshifts, and their sizes, are added into the training.

The final spectroscopic sample is the CS82 photo-$z$ training sample used to produce a photo-$z$ catalogue for the CS82 Survey. The training sample is similar to the first spectroscopic sample, but this time with the addition of quasar spectra (motivated by the results from Section~\ref{sec:res_quasar}, where the inclusion of quasar spectra in training improves the photo-$z$ quality of the catalogue). This sample is used in Section~\ref{sec:app} to produce photo-$z$ point and pdf estimates for the CS82 morphology catalogue by training $(i)$ SDSS $ugriz$ $+$ multiple morphological parameters on objects that have photometry, and $(ii)$ CS82 $i$-band $+$ multiple morphological parameters on objects that do not have SDSS photometry. Representability is solved by reweighting as described in Section~\ref{sec:reweight}, but we also ensure that the target sample has the same photometric and morphological cuts and limits as the training sample. The photometric sample selection is the same as the CS82 sample (\texttt{MASK=0}, \texttt{0$\leq$FLAGS$\leq$3}, \texttt{MAGERR\_AUTO<0.1086} and \texttt{SPREAD\_MODEL\_SER>0.008}), this results in a training set of $64\,591$ objects and a target set of $5\,777\,379$ objects in which photo-$z$'s will be estimated. Approximately $18$ per cent of objects in the target set do not have SDSS $ugriz$ photometry and thus would have their morphological redshift (morpho-z) derived using CS82 Kron $i$-band magnitude and morphology.

\begin{figure}   
\includegraphics[width=1.00\linewidth]{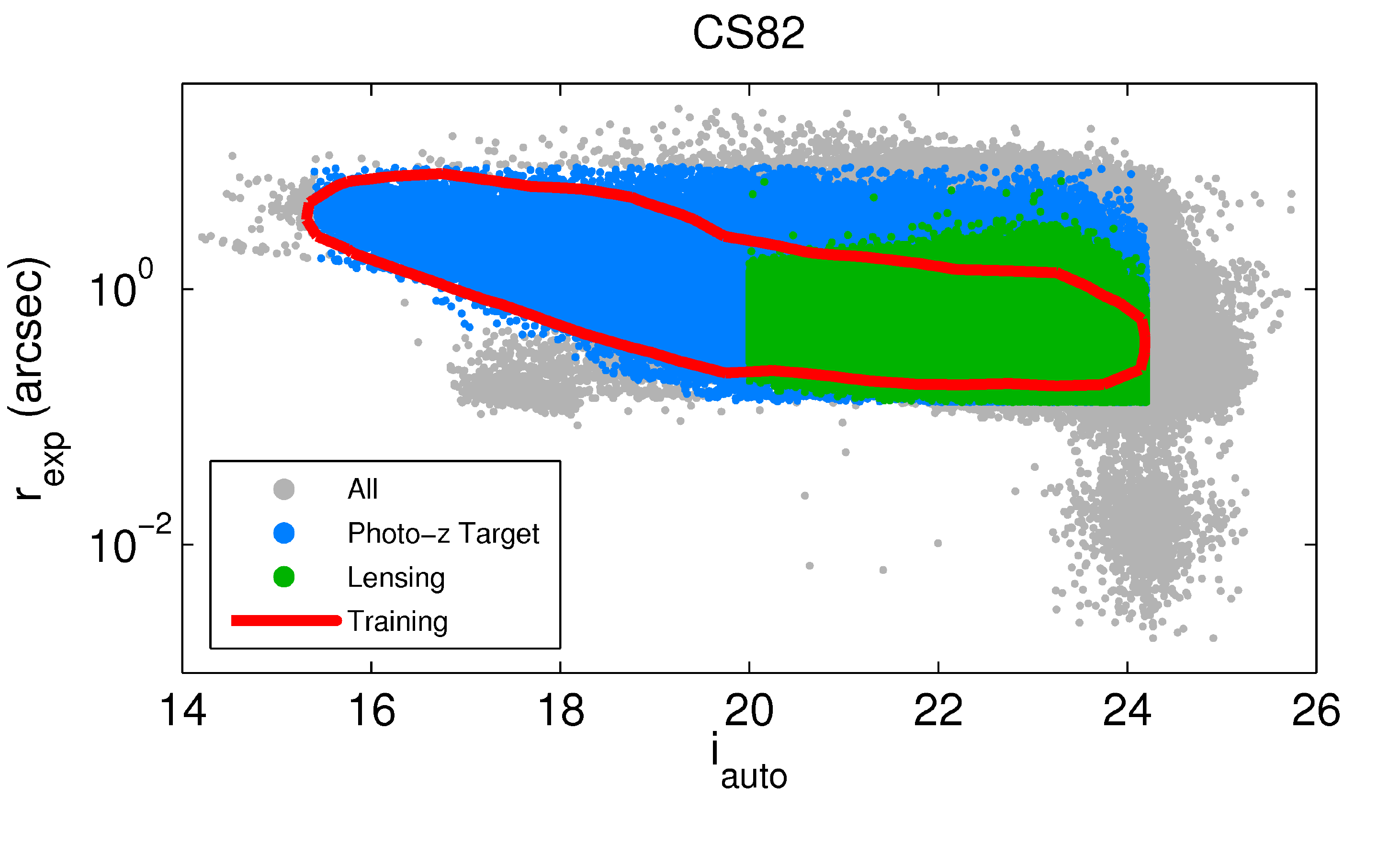}
\caption{Plot of exponential radius \textit{vs} Kron $i$-band magnitude of the CS82 general sample (red contour), compared to the CS82 photo-$z$ target sample (blue), its lensing subsample (green), and the full CS82 galaxy morphology catalogue (grey).} \label{fig:corr_target}
\end{figure}

For weak lensing analyses, a subset of the target set has been selected for further evaluation, referred to as the lensing subset. This subset has extra sample selections as follows: the objects should have good LensFit shape measurements \citep[\texttt{WEIGHT>0},][]{miller_bayesian_2013} and be classified as galaxies in FitClass\footnote{In particular, this limits the sample to $i>20$ as brighter galaxies tend to be too large to be processed efficiently by the LensFit algorithm.} (\texttt{FITCLASS=0}). This subset has $3\,536\,783$ objects. The size-magnitude diagram of the training and target sample is shown in Fig.~\ref{fig:corr_target}.

Objects in all spectroscopic samples are divided equally into three portions for training, validating and testing respectively. This size ratio between training and testing set was chosen in order to keep the number of training objects as high as possible, much higher than the training size threshold of about $2\,000$ objects suggested by \citet{collister_annz:_2004} and \citet{bonfield_photometric_2010}. A larger training size is required in our study for two reasons. Firstly, our study uses up to $10$ input parameters ($5$ more than the studies mentioned above). Secondly, our study uses reweighting, objects with high weights are rare to begin with, thus larger training sets will prevent overtraining on small numbers of high-weighted (faint and high redshift) objects. We have also tested and verified that the size ratio between the training and testing sets does little impact on results, as long as a relatively large training set is used. For instance, we find that the results using training-testing ratios between 1:1, 1:2 and 1:3 differ in root-mean-square and 68th percentile error of at most $0.5$  and $1.2$ per cent respectively when the same training set is used.

\section{Morphology and Metrics}\label{sec:method}

\subsection{Morphological parameters}\label{sec:param}
All morphological parameters used in this study are taken from the CS82 Survey. We make use of morphological parameters derived from $i$-band de Vaucouleur, exponential and S\'ersic profile fits. When reporting radius, axis-ratio and mean surface brightness, we choose the values derived from the exponential profile fit, because we find them to be more robust compared to the de Vaucouleurs' or S\'ersic fits, however the difference in training results is negligible. The morphological parameters used in this study are listed as follows:

\begin{enumerate}
\item Radius, $r_{\rm exp}$, or more specifically, the semi-major axis of the object;

\item Axis-ratio $Q=r_{A}/r_{B}$, where $r_{A}$ and $r_{B}$ are the semi-major and semi-minor axes of the object respectively. This is a form of measure for ellipticity;

\item Circularised radius, $r_{\rm c}=r_{\rm exp}\sqrt{Q}$. This is a form of measure for galaxy size (or area), which is independent of the object's ellipticity. In this study we used two different circularised radii for comparison, one from SDSS ($r_{\rm c,SDSS}$) and the other from CS82 ($r_{\rm c,CS82}$) to study if the quality of morphology affects the photo-$z$ results.

\item Mean surface brightness, $\mu$ in units of mag~arcsec$^{-2}$;

\item Shape probability $p$, which indicates if the object's shape is closer to a disc galaxy (exponential fit) or to an elliptical galaxy (de Vaucouleurs fit):
\begin{equation}
p=\frac{\chi^2_{\rm deV}}{\chi^2_{\rm deV}+\chi^2_{\rm exp}};
\end{equation}
$p$ takes values between $0$ and $1$; it compares the reduced $\chi^2$ values of both fits;

\item S\'ersic index $n$, which controls the slope of the S\'ersic profile. Setting $n=1$ reverts to an exponential profile, while $n=4$ reverts to a de Vaucouleurs profile.
\end{enumerate}

In this study we do not use the concentration index due to the absence of Petrosian fits in the CS82 morphology catalogue, and not all of the objects from SDSS Stripe-82 have valid Petrosian data. Therefore, including concentration index in our study would result in cutting the sample sizes by half, and would also introduce selection effects. Besides, it has also been shown that no significant improvements in photo-$z$'s were found when concentration index was included in the training \citep{way_new_2009,way_galaxy_2011,jones_analysis_2017}.

\begin{figure*}
\centering
\includegraphics[width=\linewidth]{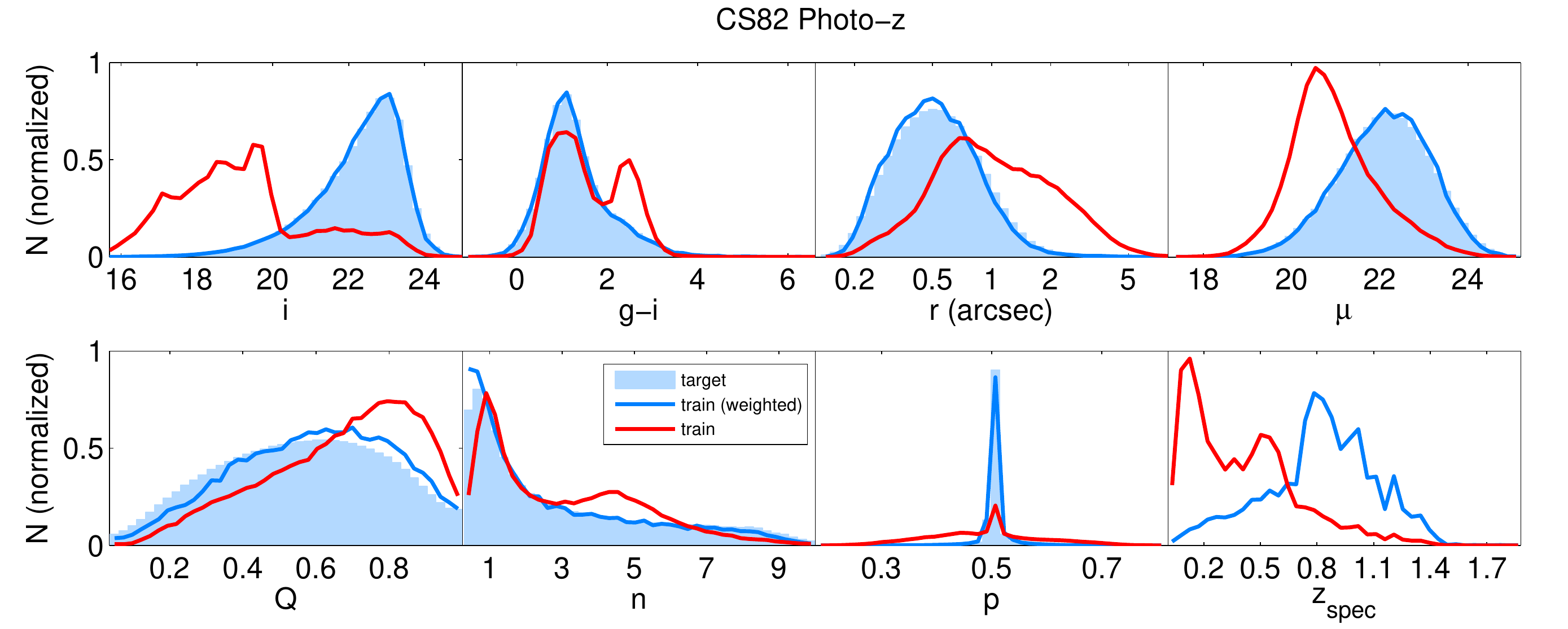}
\caption{Distribution of $i$-band magnitude, colour $g$-$i$, radius $r_{\rm exp}$, surface brightness $\mu$, axis-ratio $Q$, S\'ersic index $n$, shape probability $p$ and spectroscopic redshift for the target set (blue histogram), compared to the distribution of the training set, both weighted (blue line) and unweighted (red line). Note the reweighting is only done in terms of $i$ and $g$-$i$, but works well for all other parameters considered.} \label{fig:cs82_reweight}
\end{figure*}

As mentioned in Section~\ref{sec:reweight}, the training set is reweighted with respect to the CS82 photo-$z$ target set using the $i$-magnitude and the colour $g$-$i$. Fig.~\ref{fig:cs82_reweight} shows the distribution of $i$, $g$-$i$ and all the morphological parameters mentioned above, comparing the target set, the training set and the reweighted distribution of the training set. From this figure we confirm that the reweighting in just $i$ and $g$-$i$ is able to reweight all other morphological parameters to represent the distribution of the target set. In the final panel we also see the expected redshift distribution of the target set, which should peak around $z_{\rm spec}\approx 0.8$. We also refer the reader to Fig.~\ref{fig:corr_param} in the Appendix, which shows the correlation between these morphological parameters and spectroscopic redshift, both in weighted and unweighted densities.

\subsection{Metrics}\label{sec:metric}
The three metrics used to quantify the performance and overall distributions of the photo-$z$'s are as follows. Note that all metrics are scaled by $1+z_{\rm spec}$:

\begin{enumerate}
\item \textit{root-mean-square error} $\sigma_{\rm RMS}$, 
\begin{equation}
\sigma_{\rm RMS}=\sqrt{ \frac{\sum w_i \Delta z^2_i}{\sum w_i} }\ ,
\end{equation}
where $w_i$ is the weight of the object (obtained from the reweighting algorithm described in Section~\ref{sec:reweight}), and $\Delta z_i = (z_{\textrm{phot,}i}-z_{\textrm{spec,}i})/(1+z_{\textrm{spec,}i})$, is the difference between the photometric and spectroscopic redshift, scaled by $1+z_{\rm spec}$. Note that $\sigma_{\rm RMS}$ is calculated without outliers removed, and thus measures the overall scatter of the sample.

\item \textit{68th percentile error} $\sigma_{68}$, the half width of the weighted distribution of $\Delta z_i$ containing $68$ per cent of the objects. This measures the core width of the photo-$z$ distribution, with reduced sensitivity to outliers.

\item \textit{outlier fraction} $\eta_{\rm out}$, which is the weighted percentage of objects for which 
\begin{equation}
\left| \Delta z_i \right| \geq 0.15\ ,
\end{equation}
as introduced by \citet{ilbert_accurate_2006}. This metric identifies the percentage of objects with large outliers. The specific threshold value is chosen to enforce consistency with previous literature.


\end{enumerate}

\subsection{ANNz2 ODDS parameter}\label{sec:odds}
In this paper we introduce an \textit{ODDS parameter} (denoted by $\Theta$) for \textsc{annz2} output, originally known as the `Bayesian odds' in the template-based photo-$z$ code \textsc{bpz} \citep{benitez_bayesian_2000}:

\begin{equation}
\Theta=\int^{z_{\rm peak}+\delta z}_{z_{\rm peak}-\delta z} p\left( z|m_{j}\right) dz.
\end{equation}

$\Theta$ ranges between $0$ and $1$. It measures the probability mass between the values $z_{\rm peak}\pm \delta z$, where $\delta z=k(1+z_{\rm peak})$. $p\left( z|m_{j}\right)$ is the pdf distribution of the output photo-$z$, $m_{j}$ refers to the list of inputs used (broadband filters and morphology), and $z$ is the redshift. In this study we set $k=0.067$. This is chosen such that not too many objects end up having $\Theta=1$ (\citeauthor{benitez_bayesian_2000} used $k=3\times 0.067$ for \textsc{bpz}). Having $\Theta$ closer to $1$ implies that the $z_{\rm phot}$ obtained is more reliable. A mean ODDS value $\bar{\Theta}$ for the sample is also calculated and used in Section~\ref{sec:res_pz}.

\section{Morphological redshifts in a general sample}
\label{sec:res_general}

\label{sec:res_diff_param}

The first question we would like to address is whether the addition of morphological quantities to neural network training helps to obtain better redshifts. As discussed in the introduction, past studies have not provided a clear picture, which may at least partially be due to the details of the galaxy samples used. We selected a sample of galaxies which reaches a magnitude as faint as $i\sim 24$, close to representing current large-scale galaxy surveys like KiDS and DES, although we note that this sample does not cover the range of magnitudes expected from Stage IV surveys such as LSST. We applied reweighting on the spectroscopic samples to obtain a representative training set and train a neural network with several combinations of morphological parameters added to multi-band fluxes. More specifically, we use training sample no. $1$ as described in Section~\ref{sec:samples}. We perform several runs with input parameters $ugriz+m$, where $m$ is a single (or a set of) morphological parameter(s) from SDSS and CS82. Results are then compared to training with $ugriz$ inputs alone. Table \ref{tab:res_choiceparam} shows the metrics $\sigma_{\rm RMS}$, $\sigma_{68}$ and $\eta_{\rm out}$ of the different photo-$z$ trainings for comparison. We measure the change in percentage of these metrics with respect to the training without morphology, calculated over the full sample. Overall, we see no significant improvement. When adding the whole set of morphological parameters chosen for this study, we reach about $4$ per cent improvement in $\sigma_{\rm RMS}$ and $3$ per cent improvement in $\sigma_{\rm 68}$.

\begin{table} 
\caption{Improvement through morphology information in root-mean-square error ($\sigma_{\rm RMS}$), $68$th percentile error ($\sigma_{68}$) and outlier fraction ($\eta_{\rm out}$) for the CS82 general sample, with respect to training with only $ugriz$. The definition of these morphological parameters can be found in Section~\ref{sec:param}.} \label{tab:res_choiceparam}
\begin{tabular}{p{1.7cm}rrrrrr}
\hline
Input vars. ($ugriz+$) & $\sigma_{\rm RMS}$ & $\Delta\%$ & $\sigma_{68}$ & $\Delta\%$ & $\eta_{\rm out} (\%)$ & $\Delta\%$ \\
\hline
-               		& $0.0921$ &        & $0.0625$ &       			& $6.42$ &        \\
\hline
$r_{\rm exp}$   		& $0.0933$ & $-1.3$ & $0.0635$ & $-1.1$ 		& $6.18$ & $3.8$ \\
$r_{\rm c,SDSS}$		& $0.0925$ & $-0.5$ & $0.0609$ & $2.5$ 			& $5.95$ & \green{$7.4$} \\
$r_{\rm c,CS82}$		& $0.0924$ & $-0.3$ & $0.0611$ & $2.3$ 			& $6.31$ & $1.7$  \\
$\mu$       			& $0.0939$ & $-2.0$ & $0.0616$ & $1.4$ 			& $6.67$ & $-3.9$  \\
\hline
$Q$    					& $0.0940$ & $-2.1$ & $0.0629$ & $-0.6$ 		& $6.33$ & $1.5$  \\
$n$             		& $0.0928$ & $-0.8$ & $0.0626$ & $-0.2$ 		& $6.31$ & $1.7$ \\
$p$   					& $0.0946$ & $-2.7$ & $0.0625$ & $0.1$ 			& $6.39$ & $0.4$  \\
\hline
$r_{\rm exp},Q$			& $0.0940$ & $-2.1$ & $0.0595$ & \green{$4.7$} 	& $6.72$ & \red{$-4.6$}  \\
$r_{\rm exp},Q,\mu,n,p$	& $0.0914$ & $0.7$ 	& $0.0604$ & $3.4$			& $6.15$ & \green{$4.2$} \\
\hline
\end{tabular}
\end{table}

There are several intuitive reasons why adding morphological quantities should in principle bring improvements. First of all, as part of our morphological model-fitting, we obtain a mean surface brightness $\mu$ derived from the radius and magnitude of the model fits. We would expect surface brightness to carry redshift information since theoretically it has a $\log(1+z_{\rm spec})^{4}$ dependence with redshift. Size also correlates with redshifts through the angular diameter distance, and as we see from Fig.~\ref{fig:corr_param} in the appendix, there is indeed a correlation between size and spectroscopic redshifts, especially for brighter galaxies. We test our runs with different versions of size estimators, including radii from different model choices and circularised radii from both SDSS ($r_{\rm c,SDSS}$) and CS82 ($r_{\rm c,CS82}$). Although the CS82 survey's average seeing is half that of SDSS, this does not have significant impact on our results.

We interpret our results as stating that training with $5$ $ugriz$ bands saturates the available redshift information for a galaxy population typical of Stage-II and Stage-III optical galaxy surveys, and that morphology does not significantly help to improve photometric redshift estimation beyond this. In past investigations, it was clear that most improvements brought by morphology were seen for bright SDSS samples only. To confirm this, we trained and tested on galaxies from the SDSS main galaxy sample without any reweighting, and found that improvements as high as 13 per cent can be achieved when trained with $ugriz$ and the $5$ morphological parameters above. Therefore selection through cuts in flux, morphological parameters and indirectly through the selections of spectroscopic samples will have a strong impact on the outcome. We will indeed show in the next section that, as the availability, quality, or reliability of flux information degrades, adding morphological quantities brings quantitative and qualitative improvements to the redshift estimation process.

\section{Impact of morphology under suboptimal conditions}\label{sec:res_less_optimal}
In the previous section, we have seen that galaxy morphology has only marginal impact on photo-$z$ quality when tested in a general sample of galaxies with good $5$-band $ugriz$ photometry. In this section, we explore the possibilities of using morphology to improve photo-$z$'s in suboptimal conditions. In Section~\ref{sec:res_less_filter} we study the effects of galaxy morphology in surveys with less than $5$ broadband filters by systematically removing magnitude bands as inputs and comparing these runs with and without morphology. In Section~\ref{sec:res_bad_photometry} we study whether galaxy morphology would bring greater improvement to photo-$z$'s in surveys where the quality of photometry is low. Finally, in Section~\ref{sec:res_quasar} we assess if galaxy morphology improves the photo-$z$ quality in a situation where the separation between point sources and extended sources is imperfect. Here we once again remind the reader that the results in this section are evaluated on the testing set.

\subsection{Limited number of filters}\label{sec:res_less_filter}

It is generally accepted that $4$ or -- ideally -- $5$ photometric bands are necessary for measuring photometric redshifts with the accuracy required by the main scientific goals of modern galaxy surveys. For instance, weak lensing surveys like DES, LSST or KiDS for which coarse line-of-sight resolution is sufficient require coverage from near-UV to near-IR in at least $4$ to $5$ bands \citep{abbott_dark_2005,ivezic_lsst:_2008,de_jong_kilo-degree_2013}. This has been empirically supported in several analyses of fewer-band surveys and in extensive studies of photo-$z$ robustness under different observational conditions, thus informing the design of some of the key experiments of the coming decade. There are, nonetheless, design choices or technical issues that might constrain surveys to work with fewer bands than would be optimal. The Dark Energy Camera Legacy Survey\footnote{\url{http://legacysurvey.org/decals/}} \citep[DECaLS,][]{schlegel_dark_2015} and the Canada-France Imaging Survey\footnote{\url{http://www.cfht.hawaii.edu/Science/CFIS/}} \citep[CFIS,][]{ibata_canada-france_2017} are examples with limited filter coverage. Technical issues can also prevent the full exploitation of survey data, such as the limited depth of SDSS $z$-band due to filter and CCD inefficiencies, or the incomplete coverage in $r$ and $i$-bands of the Red-sequence Cluster Survey-2 \citep[RCS-2,][]{gilbank_red-sequence_2011} due to bad seeing.

\begin{figure}
\includegraphics[width=\columnwidth]{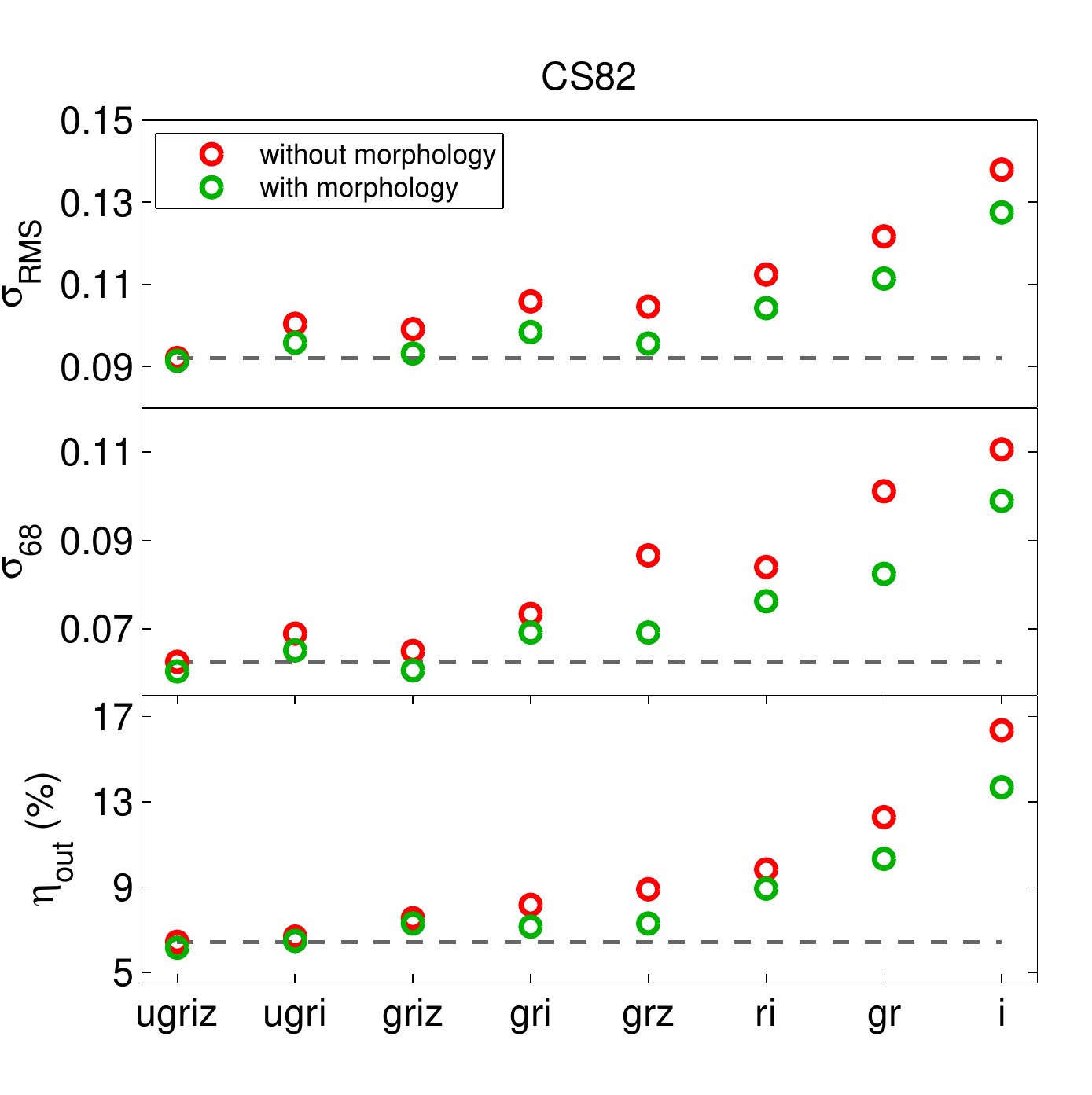}
\caption{Comparison of the root-mean-square error (top), 68th-percentile error (middle) and outlier fraction (bottom) for different photo-$z$ runs. Each panel compares pure photometry runs (red) with colour + morphology runs (green) for each combination of bands. The pure $ugriz$ run is also shown as a horizontal grey dashed line.} \label{fig:fewer_bands}
\end{figure}

Within this context, we ask whether the addition of morphology in few-band scenarios can mitigate the degradation due to the lack of detailed colour information.  We perform several \textsc{annz2} runs with different combinations of a smaller number of bands, both with and without morphology, and compare the overall performances of these runs. For morphology, we use all $5$ morphological parameters ($\mu$, $r_{\rm exp}$, $Q$, $p$ and $n$) together. Fig. \ref{fig:fewer_bands} shows the performance metrics with respect to the choice of filters used. All metrics improve with morphology relative to the photometry-only case, and more so as the number of broadband filters decreases. Taking the case where only $grz$ bands are available (similar to the case of DECaLS), we see about $14$ per cent improvement in $\sigma_{68}$ and $18$ per cent improvement in outlier fraction when morphology is included in training and reach a performance in all three metrics that is close to the full $ugriz$ case without morphology. Furthermore, a training with $1$ colour ($ri$) and morphology performs at least as well as training with $3$ filters and no morphology. In the face of these results, there is a strong case for using morphology in photo-$z$ estimation in surveys which have limited multi-band photometry, like the Red-sequence Cluster Lensing Survey \citep[RCSLenS,][]{hildebrandt_rcslens:_2016}, DECaLS and the Beijing-Arizona Sky Survey \citep[BASS,][]{zou_project_2017}.

\begin{figure*} 
\centering
\includegraphics[width=0.90\linewidth]{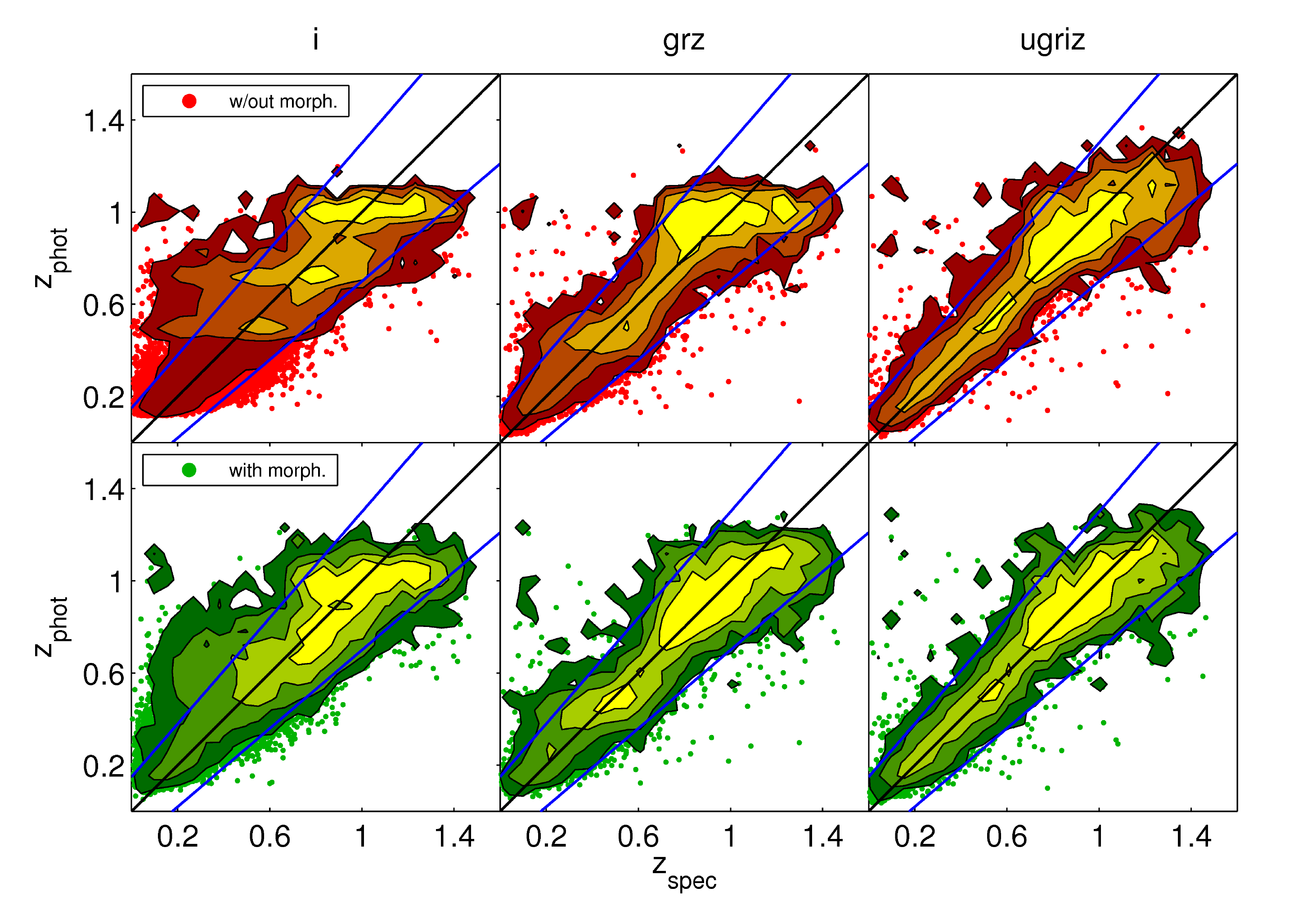}
\caption{Comparison between spectroscopic and photometric redshift based on the $i$ (left), $grz$ (middle) and $ugriz$ (right) band(s). The top panels show the training with only the respective magnitude bands, while the second row shows the training when the five morphological parameters ($r$, $Q$, $\mu$, $n$ and $p$) are included. The colours depict the weighted density of objects, and the blue lines are the limits for outliers.} \label{fig:phz_cs82}
\end{figure*}

We explore in more detail some specific qualitative aspects of particular relevance. Fig.~\ref{fig:phz_cs82} compares the results with five ($ugriz$), three ($grz$) and one ($i$) band(s). With $grz$, we see that the addition of morphology has a very noticeable impact at higher redshifts. Possibly due to biases deriving from the shallowness of the $z$-band in the SDSS survey, as shown by the contours in the figure, the neural network `saturates' after a certain redshift value, never assigning higher values; the addition of morphology redresses this high-redshift problem. Even more striking is the case with only one band $i$. As Fig.~\ref{fig:phz_cs82} shows, a single apparent magnitude provides no more than a coarse indicator for a galaxy's redshift. However, with the addition of morphological parameters, redshifts can be measured at a level of precision and accuracy that, although far from the best scenarios, makes them usable for defining broad redshift bins.

Overall, we see a robust trend where morphology provides complementary information to colours, such that the removal of colour information can be compensated by adding morphological information.

\subsection{Low-quality photometry}\label{sec:res_bad_photometry}

The quality of photometric redshifts is highly dependent on the quality of the multi-band photometry. Surveys relying on ground-based observations will inevitably accumulate data in a variety of conditions, resulting in a spatially-varying fidelity and signal-to-noise of the photometry. We will explore for an illustrative case if galaxy morphology is able to salvage the quality of photo-$z$'s in situations of poor photometry.

Stripe-82 is one of the best regions of the sky to conduct this study due to the multiple repeated scans across this region. Prior to Fall $2004$, all observational runs were taken under photometric conditions as required for imaging in the SDSS Legacy Survey \citep{york_sloan_2000}. Repeated imaging from these $84$ runs and a few later runs with seeing better than $2\arcsec$, sky brightness less than $19.5$ mag arcsec$^{-2}$ and extinction less than $0.2$ mag were processed for co-addition \citep{annis_sloan_2014}, which is the photometry used throughout this paper (and by our standards is considered `good photometry' in this section). This co-added photometry was designated \texttt{run=106,206} in the SDSS CAS, and reaches approximately $2$ magnitudes fainter than the SDSS single runs, and a median seeing of $1\farcs 1$ (compared to the usual $1\farcs 4$).

In contrast to this good photometry, runs later than Fall $2005$ on Stripe-82 were made as part of the SDSS Supernova Survey \citep{frieman_sloan_2008}, and observations were done with a higher cadence and often observed under poor seeing conditions ($\ge 2\arcsec$), bright sky, non-photometric conditions and low atmospheric transparency \citep{sako_sloan_2008}. This `bad photometry', although having photometric errors at least twice as large as those from the co-add runs, was still used in science analyses after images taken under extremely poor photometric conditions were removed, and the remaining detections subjected to a photometric calibration procedure \citep{bramich_light_2008}. This resulted in a photometry subset with larger magnitude errors, especially in the $u$-band.

We constructed a sample with `bad photometry' as follows: instead of matching our spectroscopy and CS82 morphology with the good photometry from the co-add runs ($106$ and $206$), we matched them to photometric objects without restricting from which run the object's measurements were taken. This way, objects with the same spectroscopy and morphology data have been matched to two different kinds of photometry, in one case obtained from the co-add sample, and in the other from runs with lower quality. This allows us to compare the photo-$z$ performance of the same objects under the same reweighting scheme, but with different photometric quality. Differences in magnitude limits and magnitude errors per band are provided in Table~\ref{tab:compare_goodbad}.

\begin{table}
\caption{Improvement in root-mean-square error ($\sigma_{\rm RMS}$), 68th percentile error ($\sigma_{68}$) and outlier fraction ($\eta_{\rm out}$) by morphological parameter and number of filters for the CS82 sample with low-quality photometry.} \label{tab:res_badphot}
\begin{tabular}{p{1.7cm}rrrrrr}
\hline
Input vars. ($ugriz+$) & $\sigma_{\rm RMS}$ & $\Delta\%$ & $\sigma_{68}$ & $\Delta\%$ & $\eta_{\rm out} (\%)$ & $\Delta\%$ \\
\hline
-               		& $0.1117$ &        		& $0.0892$ &       			& $12.86$ &        \\
\hline
$r_{\rm exp}$   		& $0.1203$ & $-2.2$ 		& $0.0893$ & $-0.2$ 		& $14.46$ & \red{$-12.4$} \\
$r_{\rm c,SDSS}$		& $0.1138$ & $3.3$ 			& $0.0864$ & $3.2$ 			& $12.64$ & $1.7$  \\
$r_{\rm c,CS82}$		& $0.1147$ & $2.6$ 			& $0.0868$ & $2.7$ 			& $12.85$ & $0.1$  \\
$\mu$       			& $0.1137$ & $3.4$ 			& $0.0863$ & $3.2$ 			& $12.03$ & \green{$6.5$}  \\
\hline
$Q$    					& $0.1163$ & $1.2$ 			& $0.0877$ & $1.6$ 			& $13.73$ & \red{$-6.7$}  \\
$n$             		& $0.1142$ & $3.0$ 			& $0.0856$ & \green{$4.0$} 	& $13.14$ & $-2.1$ \\
$p$   					& $0.1160$ & $1.5$ 			& $0.0863$ & $3.2$ 			& $13.27$ & $-3.2$  \\
\hline
$r_{\rm exp},Q$			& $0.1130$ & \green{$4.0$} 	& $0.0861$ & $3.5$ 			& $12.46$ & $3.1$  \\
$r_{\rm exp},Q,\mu,n,p$	& $0.1093$ & \green{$7.2$} 	& $0.0827$ & \green{$7.3$}	& $11.04$ & \green{$14.2$} \\
\hline
\end{tabular}
\end{table}

We conduct the exact same test as we did for the good photometry in Sections~\ref{sec:res_general} and~\ref{sec:res_less_filter}, adding various different morphological parameters in training while also varying the number of broadband filters used. Table~\ref{tab:res_badphot} shows the results of this run, which is a direct comparison with Table~\ref{tab:res_choiceparam}. We see a higher improvement rate in photo-$z$ in the bad photometry sample compared to the good one, especially when all $5$ morphological parameters are included, yielding a relative improvement in outlier fraction as high as $14.2$ per cent. We also see that even when low-quality morphology is used ($r_{\rm c,SDSS}$) in this case, the improvement brought is still higher than the former, although not very significant. However, it is evident that the general metrics in this sample are substantially worse than in the good photometry case, even with the help of morphology.

\begin{figure}
\includegraphics[width=\columnwidth]{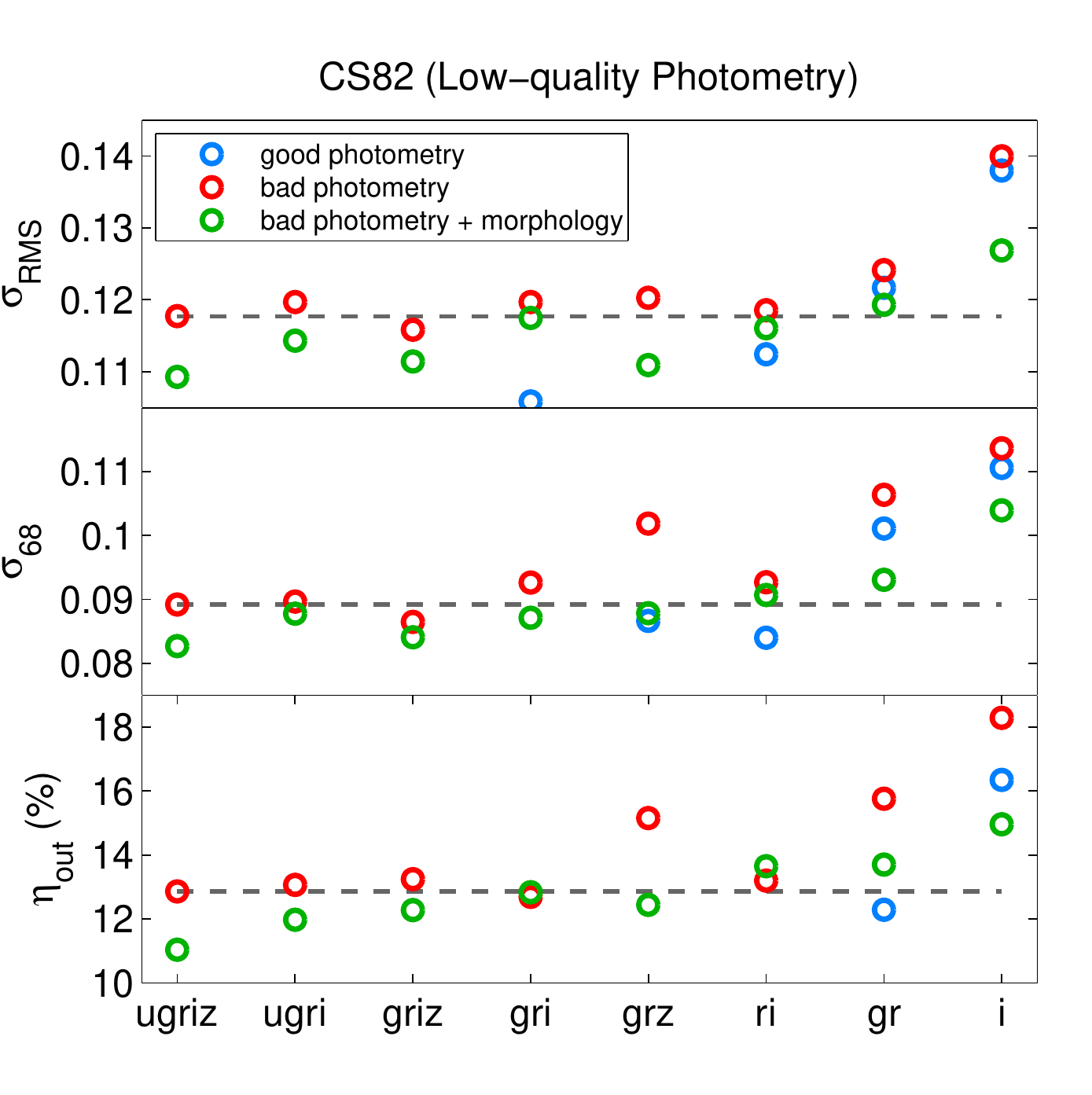}
\caption{Comparison of the root-mean-square error (top), 68th-percentile error (middle) and outlier fraction (bottom) for photo-$z$ training with different number of filters. Each panel compares pure photometry runs (red) with colour + morphology runs (green) for each combination of bands for the low-quality photometry set, compared to the pure photometry run of the good quality photometry set (blue). The pure $ugriz$ run for the bad photometry case is also shown as a horizontal grey dashed line. Improvements due to the addition of morphology are visible across the board.} \label{fig:badphot}
\end{figure}

We also assess the case when fewer filters are used with and without morphology, and the results are summarised in Fig.~\ref{fig:badphot}. Here we see that the when fewer filters are used, the improvement that morphology yields is on average larger than when good photometry is used, especially for the $ugriz$, $grz$ and $i$ cases. It is worth noting that photo-$z$'s produced with bad photometry with only $2$ bands and morphology can yield performance metrics as good as $5$ bands without morphology. We see that the metrics in this case are still far from the case when good photometry is used (blue circle), except for the cases when less than $3$ filters were used. These results further strengthen the case that morphology is a valuable addition to improve the quality of photo-$z$'s under suboptimal conditions.

\subsection{Imperfect star-galaxy separator} \label{sec:res_quasar}
Star-galaxy separation remains an ongoing problem when producing photometric redshift catalogues \citep{bundy_stripe_2015}. Current tools to separate point sources (stars and quasars) from extended objects (galaxies) include morphometric approaches, machine learning or using infrared colours \citep[see][for a comprehensive discussion]{soumagnac_star/galaxy_2015}. However, since point-source separation algorithms are not perfect, the photometric  catalogue produced will still be contaminated by a small number of stars or quasars, and the photo-$z$ estimates that go with them can be wrong if the training sample is not representative of the quasar population. 

\begin{figure*} 
\includegraphics[width=0.9\linewidth]{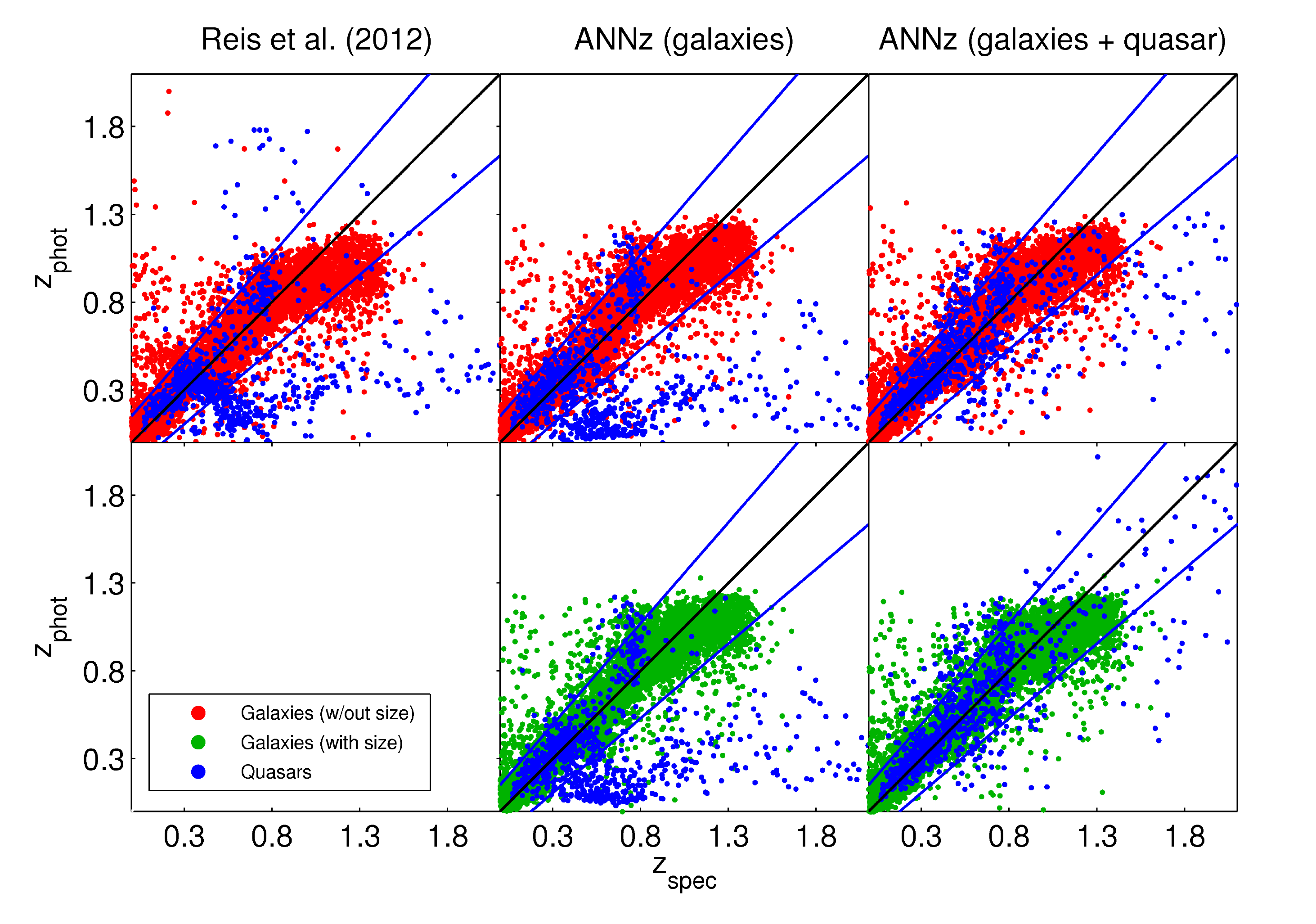}
\caption{Plot of spectroscopic \textit{vs} photometric redshift comparing the photo-$z$ produced by \citeauthor{reis_sloan_2012} (top left), \textsc{annz} training with only galaxy redshifts (middle) and \textsc{annz} training with galaxy and quasar redshifts (right). The top row shows training with only $ugriz$ as inputs (red), while the second row shows training with $ugriz$ and size (green). The joint inclusion of quasar redshifts and size in training improves the photo-$z$ for the quasars in the sample (blue).} \label{fig:phz_s82}
\end{figure*}

To study the impact of suboptimal point-source separation, the SDSS official source classification provides a good case study. SDSS uses the field \texttt{type} to separate extended objects from point sources, where \texttt{type=3} corresponds to galaxies, while \texttt{type=6} corresponds to stars/quasars. Point-source contamination in a sample of \texttt{type=3} objects in the SDSS co-add data set was estimated to be as high as $10$ per cent \citep{bundy_stripe_2015}, and the impact of these objects is seen in the official SDSS Co-add photometric redshift catalogue \citep{reis_sloan_2012}. For the purposes of our tests, we relax the point-source separation cut from the CS82 catalogue (\texttt{SPREAD\_MODEL\_SER<0.008}) and reproduce a similar spectroscopic sample to the SDSS co-add one \citep{reis_sloan_2012}. We cross-match spectroscopic redshifts of galaxies and quasars from the sources listed in Section~\ref{sec:spectroscopy} with \texttt{type=3} objects from Reis' photo-$z$ catalogue, and we plot the spectroscopic redshift \textit{vs} Reis' photo-$z$, as shown in the top left panel of Fig.~\ref{fig:phz_s82}. After the \texttt{type=3} selection, our catalogue has $3$ per cent contamination by quasars.

The blue dots in the first two columns of Fig.~\ref{fig:phz_s82} evidently show that the quasar photo-$z$'s are severely underestimated, becoming outliers in our galaxy scientific sample, even when morphological quantities are included in training in particular. If the ANN does not have high-redshift quasars for training, it will not be able to produce correct redshifts for these objects. Therefore we study the possibility of mitigating this problem by including quasar redshifts in the training, and whether size information brings additional improvements\footnote{In this section we only use SDSS morphology, due to the lack of quantities like S\'ersic index and the shape parameter in SDSS, we decided to use only one morphological parameter (size) instead of five.}. We remind the reader that by `size' we mean the circularised radius of a PSF-corrected exponential profile fit, which for quasars generally yields very small values although in some cases it is plausible that light from the host galaxy is picked up as well.

Using \textsc{annz}, Fig.~\ref{fig:phz_s82} compares four runs: the first two runs trained only with galaxies, with and without size as inputs (GAL and GAL+SIZE); the third and fourth runs include quasar redshifts in the training, with and without size (GAL+QSO and GAL+QSO+SIZE). We see that the improvement with size is only marginal if only galaxies are used for training. When quasars are added to the training, we correct the redshifts measured for quasars; however, this inclusion degrades the quality of all photo-$z$'s, increasing the $\sigma_{68}$ by approximately $3$ per cent. Our new important result is that the photo-$z$ performance is recovered when we add size to the inputs. Not only has this improved the quality of the photo-$z$ (especially quasars with $z_{\rm phot}>1.3$), we also find that with an $80$ per cent completeness cut in photo-$z$ error, approximately half of the quasars can be removed from the sample, and more high-redshift objects are kept when compared to ~\citeauthor{reis_sloan_2012}'s results. We also find the outlier fraction is reduced from $3$ per cent to $2.4$ per cent, which is a relative improvement of $20$ per cent. This result is a clear indication that the inclusion of quasar redshifts may provide a more reliable photo-$z$ for a catalogue of galaxy-like objects, especially in the case when an imperfect star/quasar-galaxy separator is used. 

While the inclusion of quasar redshifts in training improved the overall metrics for a sample of extended objects, we were also interested to know if this has particularly degraded the photo-$z$ quality of galaxies in the sample. We find that the degradation in galaxy photo-$z$'s when quasars are included in training is less than $1$ per cent across magnitude and colour, which we deem insignificant. In fact, we see that the training of GAL+QSO+SIZE performs better than GAL particularly for redder and larger galaxies. Galaxies which are large and red should generally have lower redshift, therefore size information helps to lower the overestimated photo-$z$'s for these objects. More surprisingly, we also find that the training of GAL+QSO+SIZE yields better photo-$z$'s for the \texttt{type=3} quasars compared to training QSO alone. Therefore for machine learning methods, we highly recommend the inclusion of quasar redshifts and morphology in training when estimating photo-$z$'s for galaxy samples.

Note that in this study the point-source contamination problem is only tackled partially: we improved the photo-$z$'s for the quasars, but not stars. We repeated this study by replacing quasars with stars (i.e., including star redshifts and size in training), however find that not only the photo-$z$'s of stars did not improve, the overall scatter and outlier fraction rate has degraded as well. This is mainly because star redshifts are extremely close to $0$, which introduced noise in the neural network instead. A possible extension to solve this is by using the ANN to conduct a secondary star-galaxy separation process: first output a flag to show the probability of the object being a star or quasar \citep[similar to the star-galaxy separation method by][]{soumagnac_star/galaxy_2015}, and the separated objects will undergo a second run to individually have their photo-$z$'s estimated based on their type.

\section{Impact of morphology on the pdf and the N(z)} \label{sec:res_pz}

In Sections~\ref{sec:res_general} and~\ref{sec:res_less_optimal} we have seen how morphology improves photo-$z$ point estimates produced by \textsc{annz2} using various morphological parameters in different conditions. However, the full photo-$z$ posterior distribution provides more information and is frequently used to estimate sample redshift distributions. In this section we study how galaxy morphology affects the shape of the individual pdfs and the redshift distribution $N(z)$ of the entire sample. We use the CS82 general sample for this study. We produced the $P(z)$ for each galaxy, testing with and without morphology ($r_{\rm exp}$, $\mu$, $Q$, $p$ and $n$) for cases of different numbers of passbands similar to Section~\ref{sec:res_less_filter}, with settings according to Section~\ref{sec:annz2}. The $P(z)$ for each object are summed up to produce the $N(z)$ distribution of the entire sample.

Firstly, we assessed the performance of our newly incorporated ODDS parameter $\Theta$ in \textsc{annz2}. When training with $ugriz$, we find that an $80$ per cent completeness cut in $\Theta$ retains $14.1$ per cent of $z_{\rm phot}>0.9$ objects, while just $0.2$ per cent are kept when using the same completeness cut in photo-$z$ error. The fraction of outliers is almost the same in the two cases. Therefore this shows that with $\Theta$ cuts we get to keep more objects from the higher redshift regime than with \textsc{annz2} photo-$z$ error.

\begin{figure}
\centering
\includegraphics[width=\linewidth]{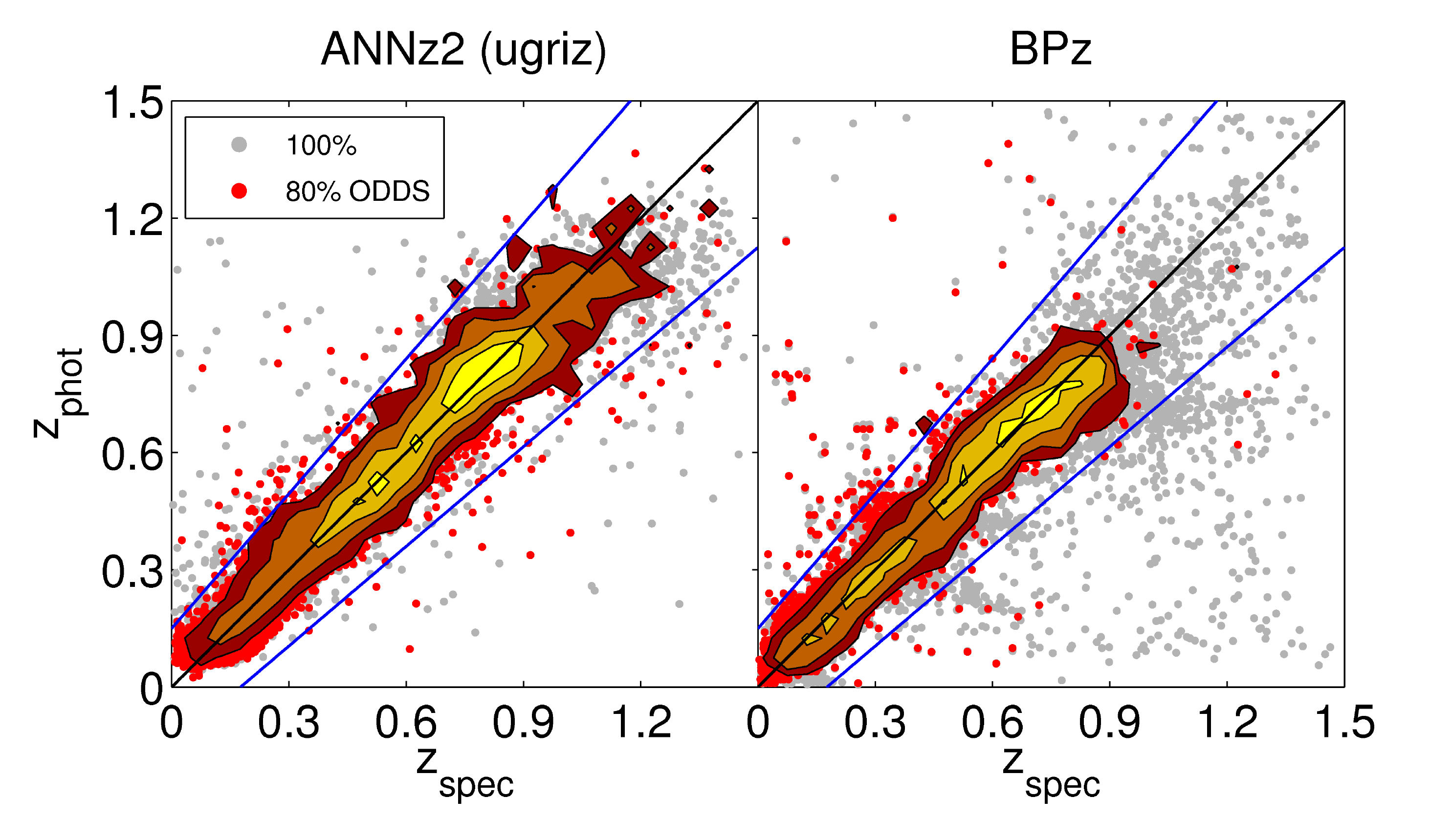}
\caption{Photometric \textit{vs} spectroscopic redshift comparing the performance of \textsc{annz2} and \textsc{bpz}, with objects with 80 per cent ODDS cut (red dots and contours) over the entire sample (grey). The contours reflect the weighted density of the objects.} \label{fig:annz_vs_bpz}
\end{figure}

We also evaluate the performance of \textsc{annz2} by comparing the photo-$z$ produced in this sample with the photo-$z$ of the same objects produced by \textsc{bpz} from the S82-MGC \citep{bundy_stripe_2015} as shown in Fig.~\ref{fig:annz_vs_bpz}. From the figure, we see that, without any cuts, \textsc{annz2} is performing better than \textsc{bpz} in terms of the number of outliers. We also applied an $80$ per cent completeness cut (discarding $20$ per cent of objects with the lowest $\Theta$) using the respective ODDS values for \textsc{annz2} and \textsc{bpz} (shaded in red), and we find that the photo-$z$'s produced by \textsc{annz2} not only efficiently reduce the number of outliers, but also keeps more objects with higher redshift ($0.9<z<1.5$), which \textsc{bpz} discarded almost completely. The results of \textsc{annz2} produce a bias at very low redshift (a visible small gap in the lower left corner of Fig.~\ref{fig:annz_vs_bpz}). This is a result of the reweighting, as the machine learning algorithm highly down-weighted bright and low-redshift objects, putting more emphasis on the higher redshift objects. The galaxies affected by this bias occupy a very small fraction of the survey volume; thus its effect is negligible across the general metrics. Besides, most objects in this low-redshift region have good spectroscopic redshifts available to be used, and they could also be removed by using a magnitude cut of $i>20$ for lensing studies.

We proceed to evaluate the effects of morphology on the $P(z)$ by studying the mean ODDS value $\bar{\Theta}$ for each run: comparing the runs with and without morphology, for the different number of filters as used in Section~\ref{sec:res_less_filter}. From a $5$ $ugriz$ band training with morphology, we find that the change in $\bar{\Theta}$ is almost negligible with morphology: $\bar{\Theta}$ increased from $0.947$ to $0.949$ when multiple morphological parameters are included, partly because $\bar{\Theta}$ is very high to begin with. The change in $\bar{\Theta}$ is however more significant when fewer filters are used, which indicates that morphology is indeed helping the ANN to improve the confidence in photo-$z$ values. We do see a high correlation between $\bar{\Theta}$ and the performance metrics ($\sigma_{\rm RMS}$, $\sigma_{68}$ and $\eta_{\rm out}$) across the number of filters used, however we see almost no correlation between the improvement in photo-$z$ and the improvement in $\Theta$ for individual galaxies when morphology is added, even when fewer filters are used. Therefore this suggests that although a high $\Theta$ does not necessarily dictate the quality of an individual galaxy's photo-$z$, it is sufficiently useful to remove outliers across an entire sample of galaxies.


\begin{figure*}
\centering
\includegraphics[width=0.8\linewidth]{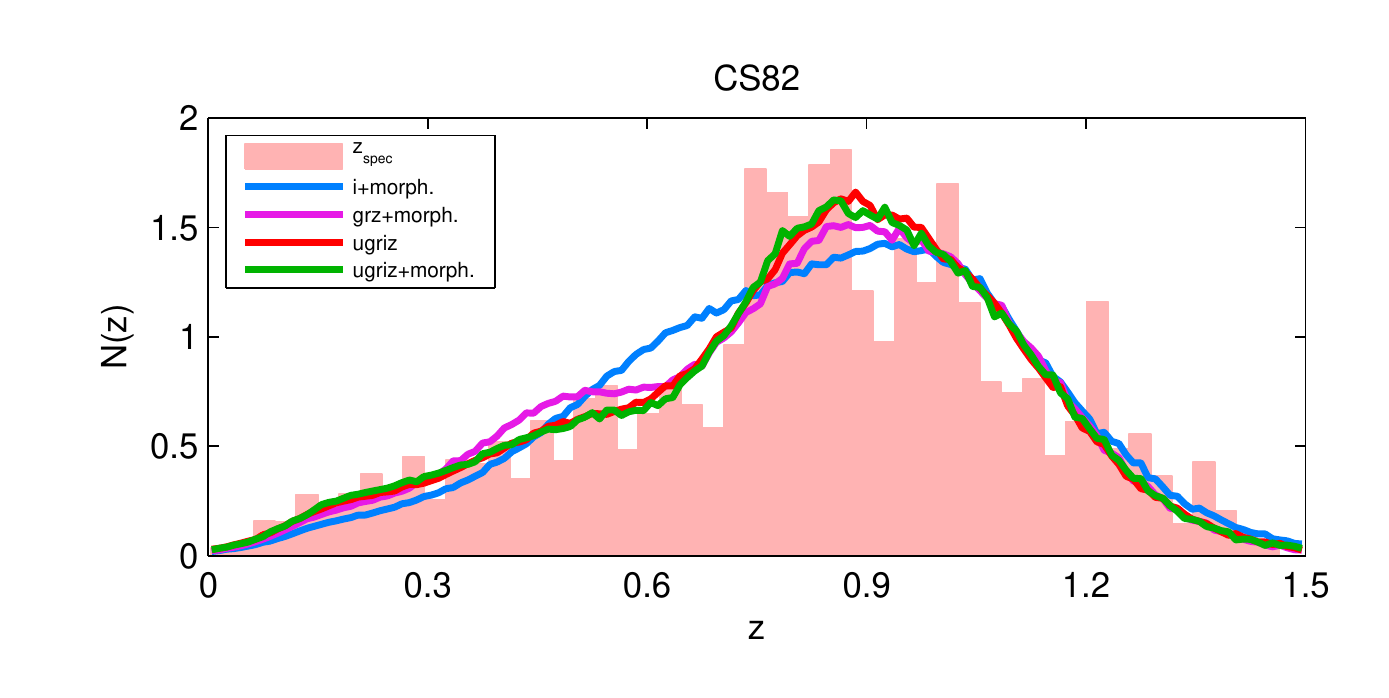}
\caption{The $N(z)$ distribution of the CS82 sample, comparing the reweighted distributions of the spectroscopic redshift (pink histogram), $i+$morphology (blue), $grz$+morphology (purple), $ugriz$ (red) and $ugriz+$morphology (green).} \label{fig:nz_pz}
\end{figure*}

Finally, we also evaluate the performance of the $N(z)$ with morphology. In this study we do not see trends or significant improvement / degradation in $N(z)$ when morphology is included in training, however we find that a relatively good $N(z)$ can be produced with at least $2$ broadband filters and morphology. In Fig.~\ref{fig:nz_pz}, we show the distribution of the weighted spectroscopic redshift (pink histogram), and compare the weighted stacked $N(z)$ for a few selected photo-$z$ runs. Here we see that the $N(z)$ produced by $ugriz$ and $ugriz$+morphology are almost indistinguishable, and resemble the overall spectroscopic distribution accurately. The $N(z)$ for $grz$+morphology is quite similar to $ugriz$ and still represents the true distribution well, while the $N(z)$ for $i$+morphology performs relatively well for $z_{\rm phot}<0.5$ and $z_{\rm phot}<1.0$. 


\section{Application: photo-$z$'s for the CS82 Morphology Catalogue}\label{sec:app}

We now utilise the results we have obtained thus far to produce a photo-$z$ catalogue for the CS82 morphology sample. As mentioned in Section~\ref{sec:cs82}, the CS82 morphology survey only has one $i$-band magnitude, and we seek to use multiple morphological parameters, SDSS $ugriz$ photometry when available, and otherwise the CS82 $i$-band magnitude, together with quasar redshifts to produce photo-$z$'s for galaxies. 

To produce this photo-$z$ catalogue, we use the 4th spectroscopic training set and its selection cuts discussed in Section~\ref{sec:samples}. We use the reweighting scheme of Section~\ref{sec:reweight} so that our training set is representative of the target sample. In this section, two redshift values are being output: the first photo-$z$ is estimated by training with $10$ inputs: the $5$ SDSS $ugriz$ magnitudes and $5$ morphological parameters ($\mu$, $r_{\rm exp}$, $Q$, $n$ and $p$). However, $18$ per cent of the objects in the catalogue do not have SDSS $ugriz$ magnitudes, and these objects make up the majority of the sample at $i>23.8$. Therefore we estimate a second redshift value called \textit{morphological redshifts} (`morpho-z'), allowing the remaining $1$ million objects in the catalogue to have redshift estimates despite not having colour information. The morpho-z will be obtained by training $6$ parameters: the CS82 Kron $i$-band magnitude\footnote{We have also included the CS82 PSF $i$-band magnitude in training, when trained together with the Kron $i$-band magnitude it would act as a proxy for the spread model of the object. However we note that the difference in results is negligible even when it is not included in training.}, $\mu$, $r_{\rm exp}$, $Q$, $n$ and $p$. Due to the lack of colour for the morpho-z case, we do the reweighting with respect to $i$ and $r_{\rm exp}$ instead, and we find it performs comparably to reweighting in $i$ and $g$-$i$, obtaining almost the exact same results as Fig.~\ref{fig:cs82_reweight}. In situations where we have multiple redshift values for each object, we provide a \texttt{ZBEST} column in the catalogue -- the `best' redshift for that object, which would be the value of either the spectroscopic redshift, photometric redshift or morphological redshift, in that order. 

\begin{table}
\caption{Comparison of photo-$z$ performance of \citet{reis_sloan_2012}, \textsc{bpz}, \textsc{eazy} and our results (\textsc{annz2}) for the testing set of the CS82 morphology photo-$z$ catalogue (the target sample). Note that these metrics have been weighted in accordance to the densities of the target sample, and no ODDS cuts have been applied.} \label{tab:cs82_phz}
\begin{tabular}{lrrr}
\hline
\bf photo-$z$ methods 	& $\sigma_{\rm RMS}$ & $\sigma_{\rm 68}$ & $\eta_{\rm out} (\%)$ \\
\hline
Reis (ANN)				&	$0.0966$	&	$0.0677$	&	$8.26$  \\
\textsc{bpz}						&	$0.1473$	&	$0.0911$	&	$19.46$ \\
\textsc{eazy}					&	$0.1234$	&	$0.0867$	&	$11.89$ \\
\hline
\textsc{annz2} ($ugriz$)			&	$0.0915$	&	$0.0606$	&	$5.97$  \\
\textsc{annz2} ($ugriz$+morph.)	&	$0.0872$	&	$0.0583$	&	$5.15$  \\
\textsc{annz2} ($i$+morph.)		&	$0.1366$	&	$0.1065$	&	$15.89$ \\
\hline
\end{tabular}
\end{table}

Point estimate and pdf of photo-$z$'s and morpho-z's are calculated, and the redshift distribution $N(z)$ is produced as well. In order to evaluate the performance of our results, we select a number of objects from the spectroscopic sample as our testing set, and we calculate the weighted performance metrics for each run without any ODDS cut, which are summarised in Table~\ref{tab:cs82_phz}. Our standard $ugriz$ photo-$z$ results outperform those by \cite{reis_sloan_2012}, \textsc{bpz} and \textsc{eazy} which we put primarily down to the reweighting. Adding morphology to $ugriz$ photometry further improves all metrics for this sample, while the morpho-z's perform similarly to \textsc{bpz} at least in these global metrics. Note that in order to compare the performance between our photo-$z$ and morpho-z results, the testing set comprises objects with colour information.

\begin{figure*}
\centering
\includegraphics[width=1\linewidth]{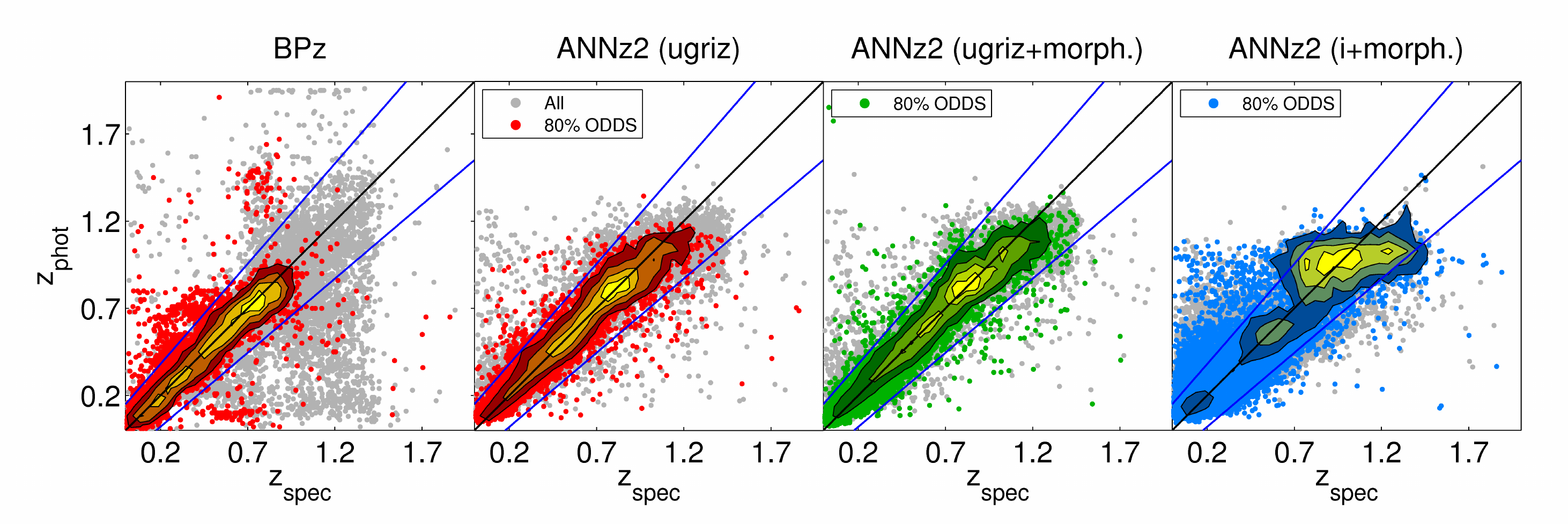}
\caption{Comparison of photometric \textit{vs} spectroscopic redshift for the CS82 photo-$z$ testing set, from left to right: \textsc{bpz}, \textsc{annz2} $ugriz$, $ugriz$+morphology and $i$+morphology. The grey points show all the objects in the testing set, while the coloured points show the $80$ per cent ODDS cut. Notably runs trained with morphology keep more of the higher redshift objects than runs trained without morphology.} \label{fig:phz_morphoz}
\end{figure*}

Fig.~\ref{fig:phz_morphoz} shows the photometric \textit{vs} spectroscopic redshift scatter plots for \textsc{bpz} and our \textsc{annz2} runs, with the contours being the weighted densities with respect to the target sample. Firstly, we see that with the correct reweighting, \textsc{annz2} successfully prioritizes to obtain better photo-$z$'s for objects between redshift $0.7<z_{\rm spec}<1.2$, which we have seen should be the expected peak redshift for the CS82 photo-$z$ catalogue in Fig.~\ref{fig:cs82_reweight}. Secondly, we also see that with an $80$ per cent ODDS cut, the training with $ugriz$+morphology outperforms the training with $ugriz$ by being able to keep more objects with higher photo-$z$'s, and this strengthens our claim that morphology is a beneficial addition to input for photo-$z$ estimation. 

\begin{figure*}
\centering
\includegraphics[width=0.9\linewidth]{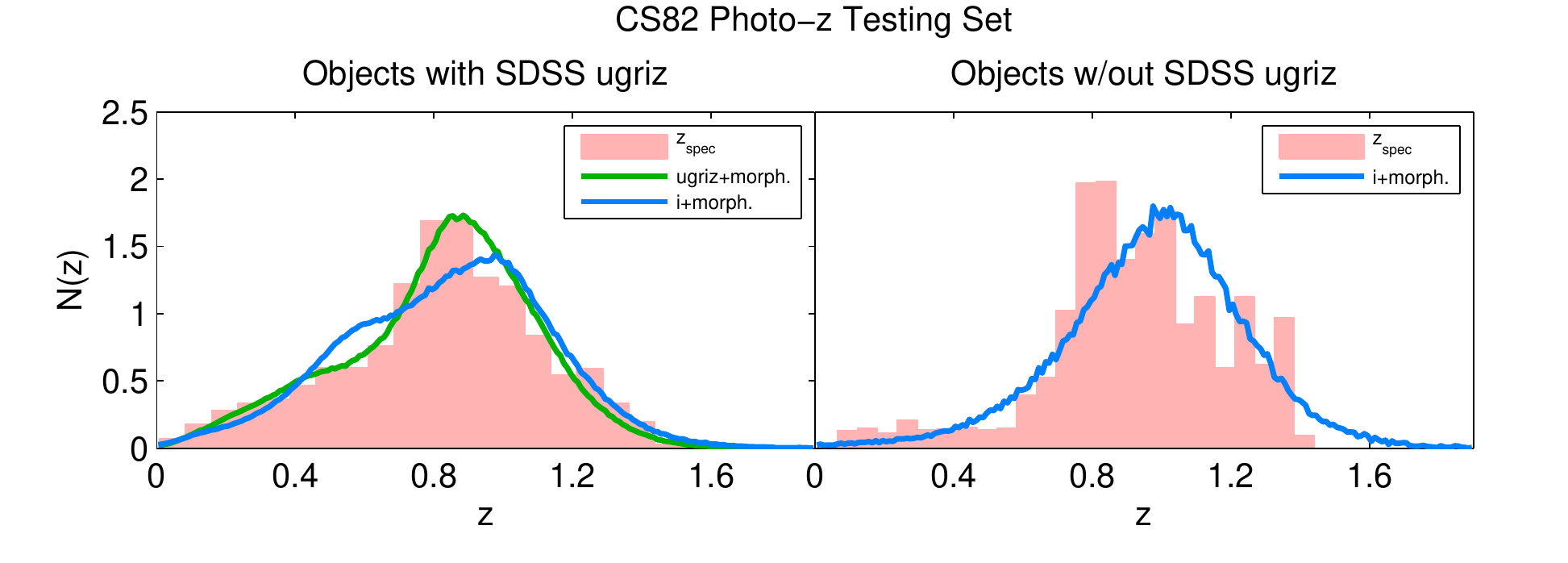}
\caption{The $N(z)$ distribution of the CS82 photo-$z$ testing set for objects with SDSS $ugriz$ magnitudes (left) and those without (right). The reweighted distributions shown are the spectroscopic redshift (pink histogram), $ugriz+$morphology (green) and $i+$morphology (blue).} \label{fig:nz_morphoz}
\end{figure*}

Finally, we evaluate the performance of $N(z)$ for each run in the testing set. In the first panel of Fig.~\ref{fig:nz_morphoz}, we find that the weighted stacked $N(z)$ produced by the $ugriz$+morphology run for objects with SDSS $ugriz$ magnitudes resembles closely the weighted spectroscopic redshift distribution. The second panel shows an $N(z)$ produced for objects which do not have $ugriz$ magnitudes (i.e., no colour information) by stacking the $P(z)$ derived from the $i$+morphology run. The stacked $N(z)$ is a fair representation of the redshift distribution of objects without colour information, reproducing accurately the number of objects at $z\lesssim 0.6$, and recovering the bulk of the distribution at $z\sim 1$.

We produce a lensing subset of the testing set by applying cuts discussed in the last paragraph of Section~\ref{sec:samples} (resulting in a magnitude range of $20<i<24.2$). We find very similar performance for the lensing subsample; see Figs.~\ref{fig:phz_morphoz_lens} and~\ref{fig:nz_morphoz_lens}. Note that the left-most panel of Fig.~\ref{fig:phz_morphoz_lens} corresponds exactly to Fig.~1 of \citet{leauthaud_lensing_2017} for direct comparison. The morpho-z's recover the redshift distribution fairly well below $z\sim 0.5$ as well as above $z\sim 1$. These conclusions hold when applying a photo-$z$ cut, e.g. at $z>0.7$ or $z>0.8$, which suggests they can be used to define a high-redshift source galaxy bin for galaxy-galaxy lensing applications, with little overlap to be expected for lens galaxy samples at $z<0.5$. 

The photo-$z$ using $ugriz$+morphology and morpho-z produced for the CS82 catalogue, together with other photometric and morphological parameters are made publicly available. The individual pdfs for each object for the $ugriz$+morphology and $i$+morphology will be distributed in separate files from the rest of the parameters. For general purposes, we recommend to use \texttt{ZBEST} as the best photometric redshift from this catalogue, an ODDS cut of $\Theta>0.5$ is also recommended to remove outliers, which yields $\eta_{\rm out}<5\%$ and retains $98$ per cent of the sample. When using purely the morpho-z values, an ODDS cut of $\Theta>0.5$ yields $\eta_{\rm out}<11\%$ and retains $33$ per cent of the sample. To estimate the $N(z)$ of the sample, we recommend the user to stack the individual pdfs of \texttt{ZPHOT} instead of \texttt{ZMORPH}. The description of the headers of table columns can be found in Appendix~\ref{appendix_header}.

\section{Conclusions} \label{sec:disc}

Starting with the first application of machine learning methods to redshift estimation, it was investigated in previous results in the literature whether the addition of morphological information could improve photometric redshift quality \citep{firth_estimating_2003,tagliaferri_neural_2003,vince_toward_2007,kurtz_-photoz:_2007,stabenau_photometric_2008,wray_new_2008,way_new_2009,lintott_galaxy_2011,singal_efficacy_2011,jones_analysis_2017}. Results varied from considerable improvement to no improvement at all. No consensus emerged, in part due to the variety of sample selections and algorithms employed in these analyses. 

The first goal of our analysis is to clarify this situation. Using well-defined photometric and spectroscopic samples and a state-of-the-art machine learning algorithm, we reach the following conclusion: with high-quality photometry in sufficient numbers of passbands (in our case five $5$ bands), adding morphology improves photometric redshifts only mildly. However, there is substantial improvement in several cases with non-optimal photometry information. We investigate surveys with fewer bands; problematic photometry due to poor observational conditions; and realistic imperfections of the point source-extended source separation. 

In the case of bad photometry, as demonstrated by the inclusion of photometry observed under bad weather / poor seeing conditions, we observed that the inclusion of morphology in training showed a higher improvement percentage in photo-$z$ quality when compared to the case of high-quality photometry. In the case of $5$ $ugriz$ bands, the inclusion of $5$ morphological parameters decreases the outlier fraction by $14.2$ per cent; while the inclusion of morphology for training with only $2$ bands yield photo-$z$'s as good as $5$ bands without morphology. 

In the case of imperfect star-galaxy separation, we demonstrate that the simultaneous addition of size information and quasar spectra to the training stage improves photometric redshift estimation of both quasars and galaxies, and in particular promotes a strong reduction of outliers in the (impure) galaxy sample. We stress that both changes are necessary: adding quasar spectra to the training degrades galaxy photometric redshifts, and it is the addition of size that redresses the situation while reducing the outlier percentage. 

In the case of fewer bands, we demonstrate significant improvements. With several combinations of $4$ or $3$ bands ($ugri$, $griz$, $gri$, $grz$), adding morphology is roughly equivalent to using $5$ high-quality $ugriz$ bands when considering standard quality metrics such as root-mean-square error, 68th percentile error and outlier fraction. Qualitative improvements -- less easily captured by the standard metrics -- are particularly interesting: as an example, in the case of $grz$ bands, pure photometry information introduces a `cut-off' at $z\sim 1.2$, limiting photo-$z$'s to be always lower than this value. Adding morphological information corrects this artificial high-redshift limit. Since target selection of high-redshift objects for DESI will employ this data set, we argue that DECaLS is a great candidate for further exploration of the ideas demonstrated in our analysis. 

In the extreme case of using only one photometric band -- in our case, $i$ -- together with morphology, we are able to estimate redshifts that, albeit far from optimal, provide a reasonable estimate of broad redshift distributions. 

Modern photometric redshift analyses work with individual and stacked probability redshift distributions instead of point estimates. We derive an ODDS parameter from the \textsc{annz2} posterior (pdf) output and demonstrate its competitiveness, in particular in retaining accurately estimated high-redshift objects. For point estimates derived from the pdfs, as well as sample redshift distributions from stacked pdfs, we confirm the earlier findings: small improvements in scatter and outlier fractions through morphology when $ugriz$ photometry is available; substantial gains if the photometry information is less complete.

We apply the insights gained to produce a new redshift catalogue for the CS82 Survey. We demonstrate that our machine learning algorithm trained with morphology and $ugriz$ magnitudes outperforms previously employed algorithms, yielding $\sigma_{68}=0.058(1+z)$ and a outlier fraction of about $5$ per cent. The overall redshift distribution of the CS82 galaxies is accurately reproduced. For a deep subsample that lacks SDSS multi-band photometry, we derive redshifts from just morphology and the CS82 single-band flux, which we argue could be of sufficient accuracy to define a high-redshift source sample for weak lensing studies.

In this study we employed estimated morphological galaxy parameters as inputs to our machine learning algorithm. This can be understood as a pre-processing step that achieves massive data compression guided by our physical understanding of how intrinsic and apparent galaxy properties evolve with redshift. With the advent of deep learning algorithms, it is possible to train directly on the pixelated galaxy images, at the price of requiring even larger training samples \citep[see][]{hoyle_tuning_2016}. It would be interesting in future to compare the performance of the two approaches on identical samples.

Photometric redshift estimation will remain a major issue also for the next generation of galaxy surveys. It may be particularly interesting for the forthcoming ESA Euclid mission to assess the use of morphological information to determine photo-$z$. Euclid will deliver very high quality morphology measurements from space-based imaging over large parts of the extragalactic sky. Optical photometry will be provided from the ground, so that both its provenance and quality will inevitably vary across the sky and across passbands. Morphological information could be a valuable complement to mitigate against the effects of calibration or quality issues in the colours. We caution that, at the precision level of Stage-IV surveys, more work is required to study the potential correlations between redshift estimates, especially when morphology is included, and shear estimates used in weak gravitational lensing applications.

The photo-$z$, morpho-z and their respective pdfs produced for the CS82 catalogue are publicly available at \url{ftp://ftp.star.ucl.ac.uk/johnsyh/cs82/}, and the catalogue will be incorporated into the official CS82 website in the future.




\section*{Acknowledgements}
J.S. would like to thank Iftach Sadeh and Antonella Palmese for fruitful discussions on the use of \textsc{annz2} algorithm; Boris Leistedt, Stephanie Jouvel and Hendrik Hildebrandt for insights on photometric redshift representations. J.S. would also like to thank participants of the LSST photo-$z$ Workshop, Pittsburgh for feedback on the early results of this paper. B.M. would like to thank Filipe Abdalla, Emmanuel Bertin and Boris Leistedt for fruitful discussions.

J.S. acknowledges support from the MyBrainSc Scholarship, provided by the Ministry of Education, Malaysia. BJ acknowledges support from an STFC Ernest Rutherford Fellowship (grant reference ST/J004421/1). OL acknowledges support from a European Research Council Advanced Grant (FP7/291329). AC acknowledges support by the Brazilian Science Without Borders program, managed by the Coordena\c{c}\~{a}o de Aperfei\c{c}oamento de Pessoal de N\'{i}vel Superior (CAPES) fundation, and the Conselho Nacional de Desenvolvimento Cientif\'{i}co e Tecnol\'{o}gico (CNPq) agency, Fora Temer (FT). MM acknowledges support from CNPq, FT.

Based on observations obtained with MegaPrime/MegaCam, a joint project of CFHT and CEA/DAPNIA, at the Canada-France-Hawaii Telescope (CFHT), which is operated by the National Research Council (NRC) of Canada, the Institut National des Science de l'Univers of the Centre National de la Recherche Scientifique (CNRS) of France, and the University of Hawaii. The Brazilian partnership on CFHT is managed by the Laborat\'orio Nacional de Astrof\'isica (LNA). We thank the support of the Laborat\'orio Interinstitucional de e-Astronomia (LIneA). We thank the CFHTLenS team.

Funding for the SDSS-I/II/III has been provided by the Alfred P. Sloan Foundation, the Participating Institutions, the National Science Foundation, the U.S. Department of Energy, the National Aeronautics and Space Administration, the Japanese Monbukagakusho, the Max Planck Society, and the Higher Education Funding Council for England. The SDSS-I/II web site is \url{http://www.sdss.org/}, while the SDSS-III web site is \url{http://www.sdss3.org/}.

SDSS-I/II/III is managed by the Astrophysical Research Consortium for the Participating Institutions. The Participating Institutions include the American Museum of Natural History, Astrophysical Institute Potsdam, University of Basel, University of Cambridge, Case Western Reserve University, University of Chicago, Drexel University, Fermilab, the Institute for Advanced Study, the Japan Participation Group, Johns Hopkins University, the Joint Institute for Nuclear Astrophysics, the Kavli Institute for Particle Astrophysics and Cosmology, the Korean Scientist Group, the Chinese Academy of Sciences (LAMOST), Los Alamos National Laboratory, the Max-Planck-Institute for Astronomy (MPIA), the Max-Planck-Institute for Astrophysics (MPA), New Mexico State University, Ohio State University, University of Pittsburgh, University of Portsmouth, Princeton University, the United States Naval Observatory, the University of Washington, University of Arizona, the Brazilian Participation Group, Brookhaven National Laboratory, Carnegie Mellon University, University of Florida, the French Participation Group, the German Participation Group, Harvard University, the Instituto de Astrofisica de Canarias, the Michigan State/Notre Dame/JINA Participation Group, Lawrence Berkeley National Laboratory, Max Planck Institute for Extraterrestrial Physics, New York University, Pennsylvania State University, the Spanish Participation Group, University of Tokyo, University of Utah, Vanderbilt University, University of Virginia and Yale University. 

Funding for the DEEP2 Galaxy Redshift Survey has been provided by NSF grants AST-95-09298, AST-0071048, AST-0507428, and AST-0507483 as well as NASA LTSA grant NNG04GC89G.

This research uses data from the VIMOS VLT Deep Survey, obtained from the VVDS database operated by Cesam, Laboratoire d'Astrophysique de Marseille, France.

This research also uses data from WiggleZ which received financial support from The Australian Research Council, Swinburne University of Technology, The University of Queensland, the Anglo-Australian Observatory, and The Gregg Thompson Dark Energy.

\bibliographystyle{mnras}
\bibliography{Paper_1} 

\appendix

\section{CS82 photo-$z$ Catalogue Headers} \label{appendix_header}
Table~\ref{tab:catalogue} lists down all the headers and their respective descriptions for the CS82 photo-$z$ catalogue. The file \texttt{cs82\_phz.fits} has all information below except for the last one (the $P(z)$). The files \texttt{cs82\_pz\_phot.fits} and \texttt{cs82\_pz\_morph.fits} contain the $P(z)$ for each object, based on the training with $ugriz$+morphology and $i$+morphology respectively. Values that are not available are left blank in the catalogue.

\begin{table*}
\caption{List of headers and their descriptions for the CS82 photo-$z$ catalogue.} \label{tab:catalogue}
\begin{tabular}{>{\tt}ll}
\hline
\bf Col. name 	& \bf description \\
\hline 
OBJID\_CS82 	& CS82 object ID 	\\
RA				& right ascension ($\deg$)	\\
DEC				& declination ($\deg$)		\\
FLAGS			& source extraction quality flag (ranges from $0$ to $3$, with $0$ being the best) \\
WEIGHT			& lensFit shape measurement flag. \texttt{WEIGHT} $>0$ indicates good shape measurement	\\
FITCLASS		& lensFit star-galaxy classifier: $1=$ stars; $0=$ galaxies; \\
				& $-1=$ no usable data; $-2=$ blended objects; $-3=$ miscellaneous reasons; $-4=\chi^2$ exceeded critical value  	\\
MAG\_AUTO		& CS82 Kron $i$-band magnitude	\\
MAGERR\_AUTO	& CS82 Kron $i$-band magnitude error. Signal-to-noise ratio is \texttt{1.086/MAGERR\_AUTO}	\\
MAG\_EXP		& CS82 exponential fit $i$-band magnitude	\\
MAG\_PSF		& CS82 PSF $i$-band magnitude	\\
REFF\_EXP		& exponential fit effective radius (arcsec)	\\
ASPECT\_EXP		& exponential fit axis-ratio	\\
MU\_MEAN\_EXP	& exponential fit mean surface brightness	\\
P\_EXP			& exponential shape probability: $\sim 1$ $\rightarrow$ disc galaxy; $\sim 0$ $\rightarrow$ elliptical galaxy	\\
N\_SER			& S\'ersic index	\\
SPREAD\_MODEL\_SER		& S\'ersic spread model, the star-galaxy separator we used for this study. All objects in this catalogue \\
				&	have SPREAD\_MODEL\_SER $>0.008$, which are considered extended objects (galaxies) \\
LENS        	& lensing tag. Objects with \texttt{LENS} $=1$ are objects from the lensing subsample (\texttt{FITCLASS} $=0$, \texttt{WEIGHT} $>0$) \\
OBJID\_SDSS		& SDSS object ID for objects with matched $ugriz$ broadband magnitudes (if available) \\
MAG\_DERED\_U	& SDSS dereddened $u$-band magnitude (if available)	\\
MAG\_DERED\_G	& SDSS dereddened $g$-band magnitude (if available)	\\
MAG\_DERED\_R	& SDSS dereddened $r$-band magnitude (if available)	\\
MAG\_DERED\_I	& SDSS dereddened $i$-band magnitude (if available)	\\
MAG\_DERED\_Z	& SDSS dereddened $z$-band magnitude (if available)	\\
ZSPEC	    	& spectroscopic redshift (if available) \\
ZPHOT			& photometric redshift estimated using inputs $ugriz$+morphology (if available)	\\
zMORPH			& photometric redshift estimated using inputs $i$+morphology					\\
zBEST			& best redshift for this object, in order of priority: \texttt{ZSPEC}, \texttt{ZPHOT}, \texttt{ZMORPH} \\
ODDS\_PHOT		& ODDS value for \texttt{ZPHOT} \\
ODDS\_MORPH		& ODDS value for \texttt{ZMORPH}	\\
ODDS\_BEST   	& ODDS value for \texttt{ZBEST}  ($=1$ if \texttt{ZSPEC} is used) \\
SOURCE\_SPEC	& source of spectroscopy: SDSS, BOSS, DEEP2, WIGGLEZ or VVDS (if available)	\\
CLASS\_SPEC   	& class of object based on spectral fit: GALAXY or QSO (if available) \\
z\_0 - z\_189	& $P(z)$ values for \texttt{ZPHOT} / \texttt{ZMORPH}, ranging from $z=0.005$ to $z=1.895$ in an equal step size of $0.01$	\\
\hline
\end{tabular}
\end{table*}

\section{Additional Plots and Tables}
This section includes extra figures and tables referenced in Sections~\ref{sec:param} and~\ref{sec:res_bad_photometry}.

\begin{figure*}
\centering
\includegraphics[width=0.40\linewidth]{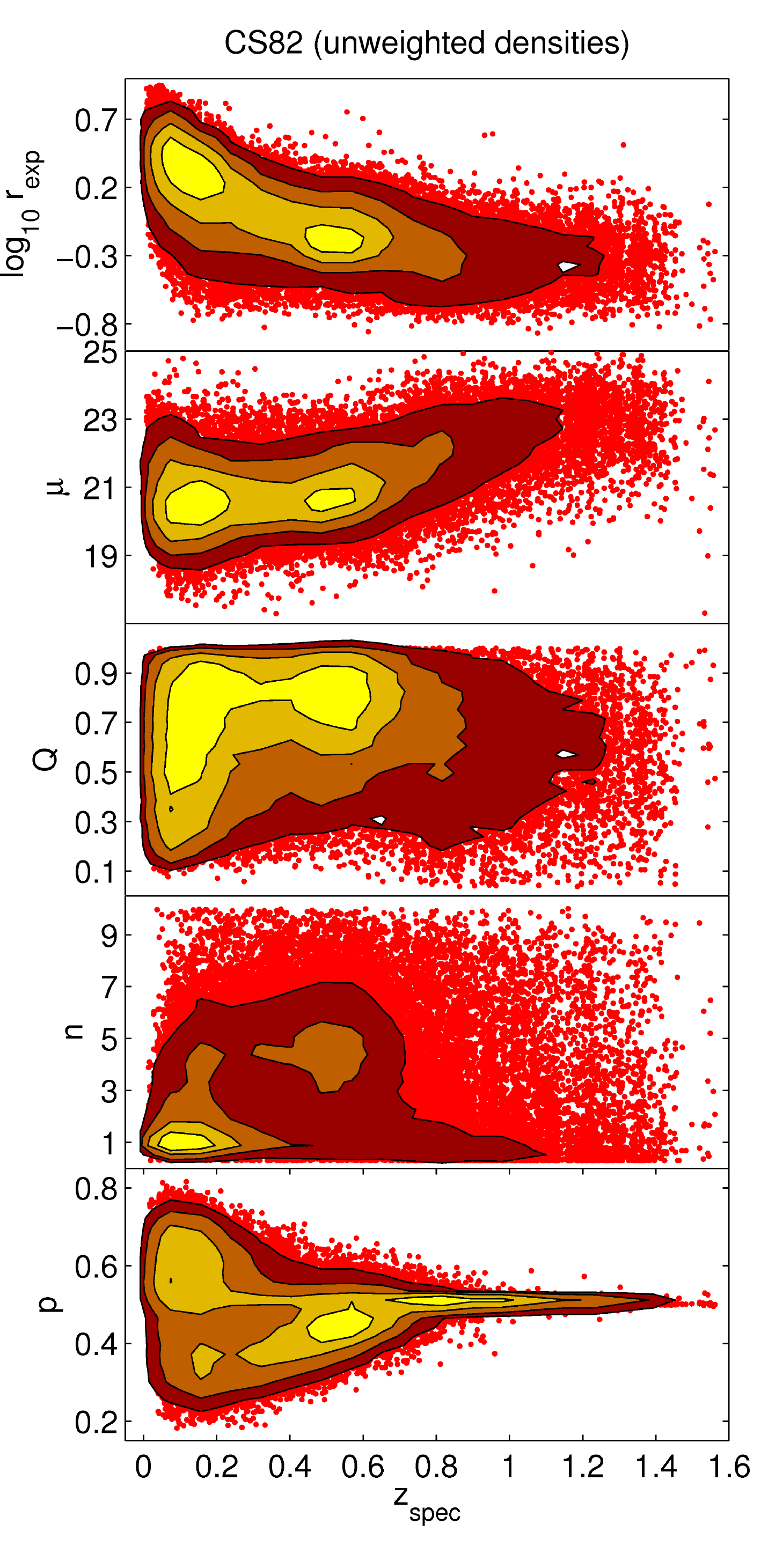}
\includegraphics[width=0.40\linewidth]{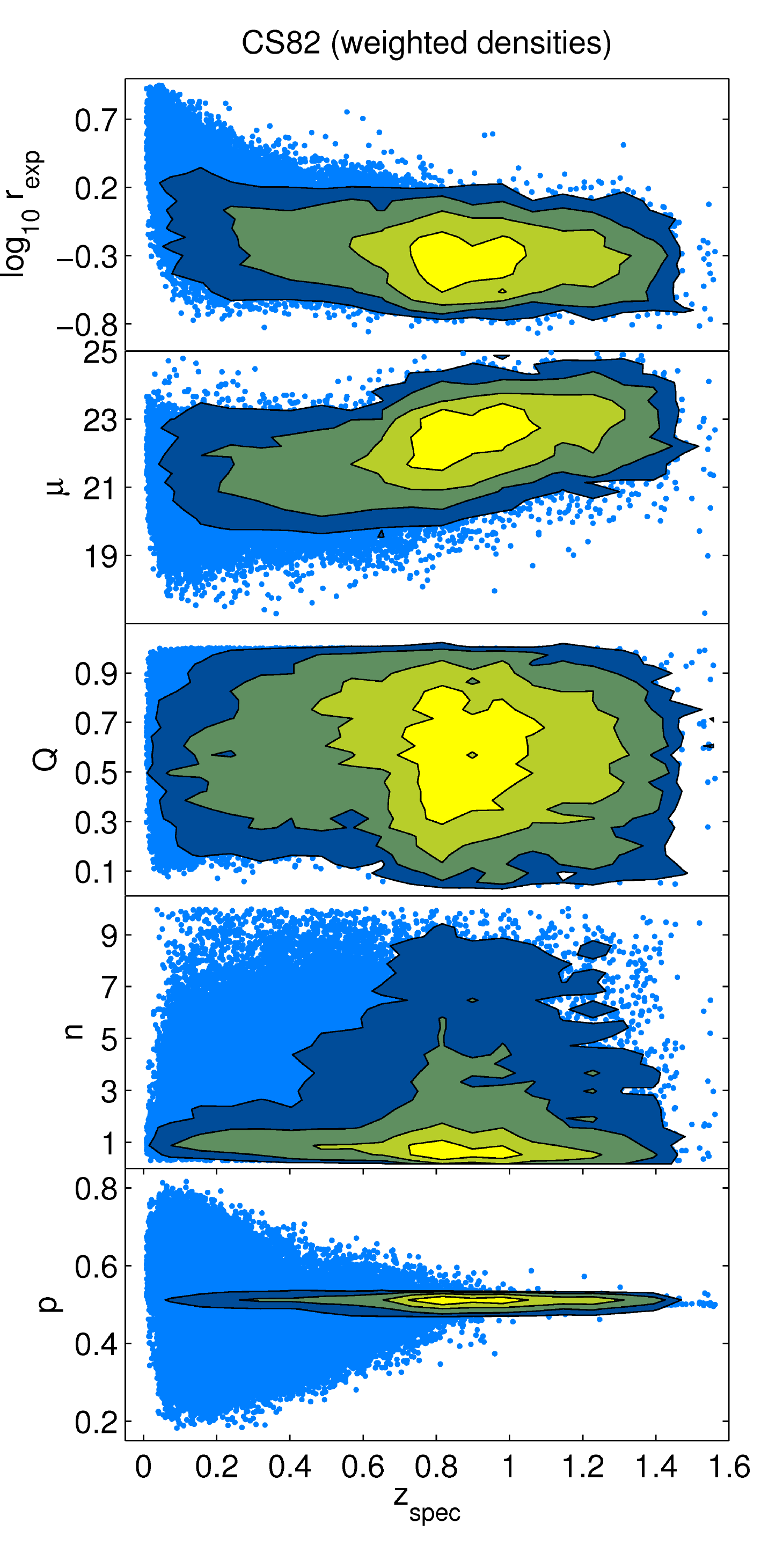}
\caption{Scatter plots of log-radius ($\log_{10} r_{\rm exp}$), surface brightness ($\mu$), axis-ratio ($Q$), S\'ersic index ($n$) and shape probability ($p$) against spectroscopic redshift for objects in the CS82 photo-$z$ training set. The contours represent the unweighted (red, left) and weighted (blue, right) density of objects, weighted with respect to the CS82 photo-$z$ target sample.} \label{fig:corr_param}
\end{figure*}

\begin{table}
\caption{Comparison between the good and bad photometry samples. Top: The mean absolute difference in magnitude for the respective bands. Bottom: Comparison of the mean magnitude errors for the respective bands for the good and bad photometry samples. It can be seen that the difference in $u$-band magnitude is much higher than other bands, and the bad photometry sample has mean magnitude errors $2$ to $3$ times larger than those of the good photometry sample.} \label{tab:compare_goodbad} 
\begin{tabular}{>{\bf}l>{\bf}r>{\bf}r>{\bf}r}
\hline
Filter 	& Mean 			& Mean		&	Mean	\\
		& absolute		& error 	&	error	\\
        & difference	& (good)	&	(bad)	\\
\hline
$u$		&	$0.777$		&	$0.447$	&	$0.762$	\\
$g$		&	$0.185$		&	$0.031$	&	$0.135$	\\
$r$		&	$0.139$		&	$0.017$	&	$0.061$ \\
$i$		&	$0.134$		&	$0.015$	&	$0.049$	\\
$z$		&	$0.172$		&	$0.038$	&	$0.128$	\\
\hline
\end{tabular}
\end{table}

\section{Extra Plots for photo-$z$ Lensing Subset}
This section includes figures related to the lensing subsample of the CS82 photo-$z$ catalogue discussed in Section~\ref{sec:app}.

\begin{figure*}
\centering
\includegraphics[width=1\linewidth]{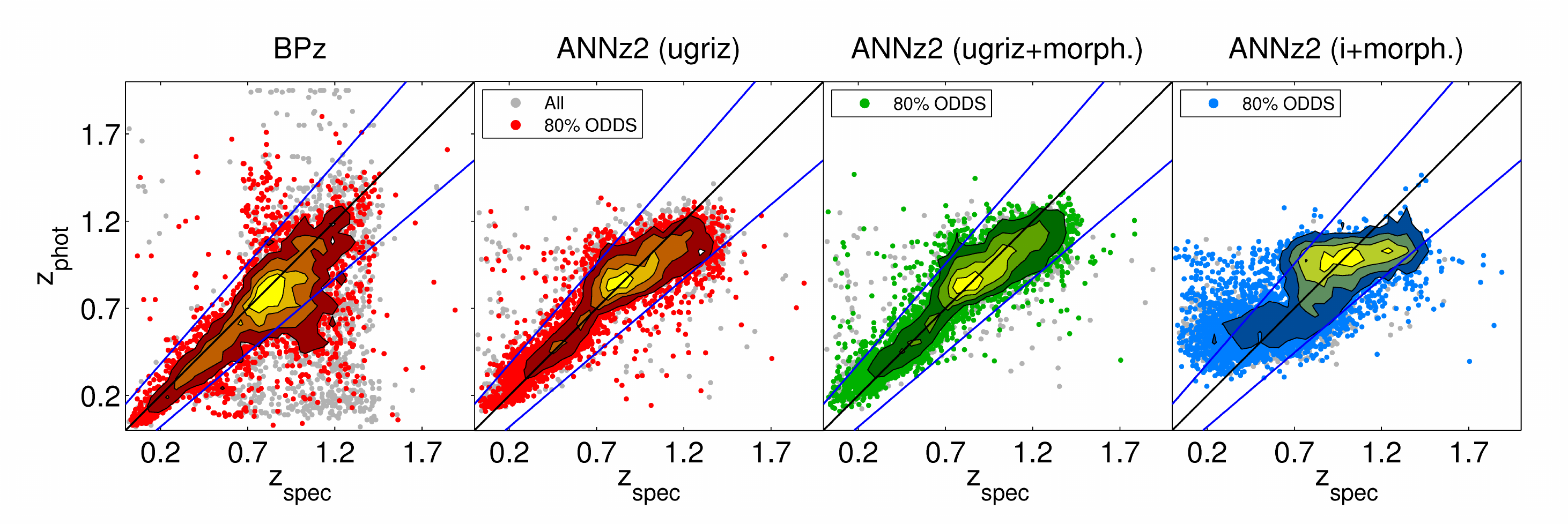}
\caption{Same as Fig.~\ref{fig:phz_morphoz}, but for the lensing subsample. The first panel on the left corresponds exactly to Fig.~1 of \citet{leauthaud_lensing_2017} for direct comparison.} \label{fig:phz_morphoz_lens}
\end{figure*}

\begin{figure*}
\centering
\includegraphics[width=0.9\linewidth]{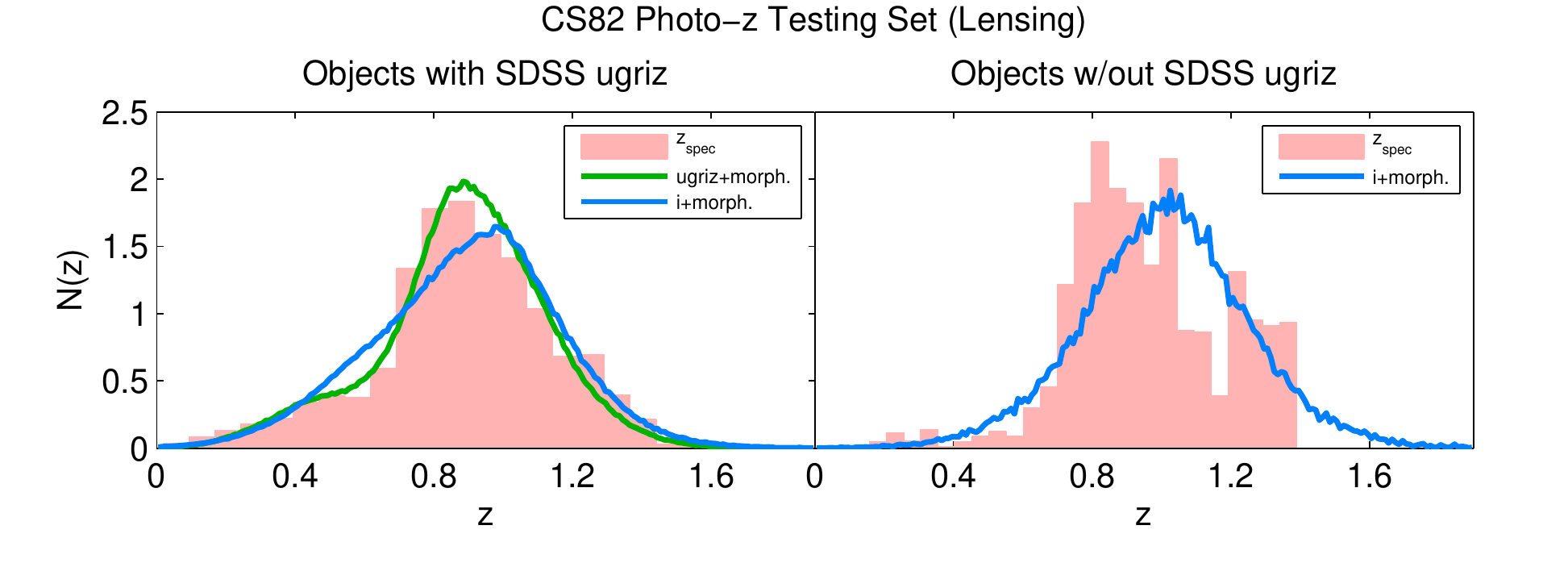}
\caption{Same as Fig.~\ref{fig:nz_morphoz}, but for the lensing subsample.} \label{fig:nz_morphoz_lens}
\end{figure*}

\bsp	
\label{lastpage}
\end{document}